\documentclass[aip,preprint]{revtex4-2}
\usepackage[english]{babel}
\usepackage{amsmath,amssymb,amsfonts}
\usepackage{graphicx}
\usepackage{mathrsfs}
\graphicspath{{figures/}}

\begin{document}

\author{\firstname{Petr} \surname{Brandyshev}}
\affiliation{Laboratory of Computational Physics, HSE University, Tallinskaya st. 34, 123458 Moscow, Russia}
\affiliation{School of Applied Mathematics, HSE University, Tallinskaya st. 34, 123458 Moscow, Russia}
\author{\firstname{Maksim} \surname{Koniushenko}}
\affiliation{School of Applied Mathematics, HSE University, Tallinskaya st. 34, 123458 Moscow, Russia}
\author{\firstname{Yury} \surname{Budkov}}
\email[]{ybudkov@hse.ru}
\affiliation{Laboratory of Computational Physics, HSE University, Tallinskaya st. 34, 123458 Moscow, Russia}
\affiliation{School of Applied Mathematics, HSE University, Tallinskaya st. 34, 123458 Moscow, Russia}
\affiliation{Laboratory of Multiscale Modeling of Molecular Systems, G.A. Krestov Institute of Solution Chemistry of the Russian Academy of Sciences, 153045, Akademicheskaya st. 1, Ivanovo, Russia}

\title{Casimir force screening by a quantum Boltzmann plasma}

\begin{abstract}
We develop a finite-temperature field-theoretical description of the Casimir
interaction between two perfectly conducting boundaries with the intervening
gap filled by a two-component quantum Boltzmann plasma embedded in a homogeneous
dielectric background. The quantum motion of the charged particles is represented by
imaginary-time worldlines, while the electromagnetic field is treated in the
Gaussian approximation. The resulting partition function contains distinct
longitudinal and transverse response contributions and an additional
quantum-induced source term with no classical counterpart. In the classical
limit the theory reduces to the usual Casimir-screening picture: the static
longitudinal mode is Debye screened, whereas the transverse electromagnetic
sector is unaffected by classical mobile charges. Finite thermal de Broglie
wavelengths modify the nonzero Matsubara response and produce the leading
quantum correction to the force. In the static long-wavelength limit, the
longitudinal dielectric function has, through the leading quantum correction,
the same expansion structure as the Lindhard response of a degenerate electron
gas under the correspondence
$k_{TF}\leftrightarrow\kappa_D$ and
$k_F\leftrightarrow\lambda_{dB}^{-1}$, while the static transverse response
reproduces the leading high-temperature Landau orbital diamagnetic
susceptibility. We investigate the crossover between classical and quantum
plasma behavior in a regime where these corrections become appreciable while
the plasma remains approximately nondegenerate and can still be treated, at
leading order, within Maxwell--Boltzmann statistics. We illustrate this crossover for
a weakly coupled electron--proton plasma and an effective
$\mathrm{Li}^{+}$--solvated-electron plasma in liquid ammonia. The latter
yields a sub-percent quantum correction to the force gradient for the
parameters considered.
\end{abstract}

\maketitle
\section{Introduction}
The Casimir interaction in an ionic medium differs qualitatively from that in
vacuum or in a passive dielectric because the electromagnetic field coexists
with mobile charged particles. Thermal charge fluctuations generate an
electrostatic contribution to the interaction and, at sufficiently large
separations, screen its longitudinal zero-frequency part. The interplay
between dispersion forces and ionic screening was already discussed in the
early studies of electrolyte films~\cite{barouch1973interdependence,gorelkin1973calculation,barnes1975statistical}. The assumption that the equilibrium fluctuations of the mobile charges can be separated from the transverse electromagnetic fluctuations was later formulated as the \textit{separation hypothesis}~\cite{jancovici2004screening}. Microscopic statistical-mechanical analyses by Jancovici and \v{S}amaj~\cite{jancovici2004screening,jancovici2005casimir} and by Buenzli and Martin~\cite{buenzli2005microscopic,buenzli2008microscopic} established the large-distance structure of the thermal Casimir force in systems containing classical mobile charges.

A complementary scattering formulation was developed by Maia Neto \textit{et al.}~\cite{maianeto2019scattering}, who explicitly included the nonlocal response of an electrolyte and the coupling between longitudinal and transverse electromagnetic channels. Their analysis showed that the zero-frequency longitudinal contribution is screened on the Debye scale, whereas the transverse-magnetic zero-frequency contribution remains unscreened. For ordinary electrolytes, the nonzero Matsubara terms are essentially unchanged by the ions. This unscreened thermal contribution was subsequently probed experimentally using optical tweezers~\cite{pires2021probing}.

In our recent work~\cite{budkovBrandyshev2025PRE}, based on the finite-temperature field-theoretical formulation developed in Ref.~\cite{brandyshevBudkov2025JMP}, we reconsidered the same problem from first principles for classical mobile charges confined between perfectly conducting plates. In the Gaussian approximation, the ionic and transverse electromagnetic contributions to the partition function separate. This result provides a microscopic justification of the separation hypothesis and is directly connected with the Bohr--van Leeuwen theorem: classical charged particles in thermal equilibrium do not modify the transverse electromagnetic fluctuations. Consequently, the classical plasma screens one of the zero-frequency sectors, while the nonzero electromagnetic Matsubara modes remain unchanged.

The situation changes when the mobile charges are treated quantum mechanically. A quantum particle is represented by an extended imaginary-time trajectory, and its finite spatial extension generates both charge-density and current fluctuations. The transverse electromagnetic field can therefore couple to the matter degrees of freedom even in thermal equilibrium. The classical separation is no longer exact, the nonzero Matsubara modes acquire a plasma contribution, and the worldline structure generates an additional fluctuation-induced source term.

The aim of the present work is to investigate the Casimir interaction in the crossover regime between classical and quantum plasma behavior, where quantum effects become appreciable while the plasma remains approximately nondegenerate and can still be described by Maxwell--Boltzmann statistics. This regime is distinct both from the strictly classical limit and from a degenerate electron gas. We derive the longitudinal and transverse response functions of a quantum Boltzmann plasma, determine the corresponding Casimir free energy, and identify the source contribution generated by the finite imaginary-time extension of the charged particles. As analytical benchmarks, we show that the static long-wavelength longitudinal response has, through the leading quantum correction, the same expansion structure as the Lindhard result under the formal replacements $k_{\rm TF}\leftrightarrow\kappa_D$ and $k_F\leftrightarrow\lambda_{dB}^{-1}$, whereas the static transverse response reproduces the high-temperature Landau orbital diamagnetic susceptibility~\cite{Lindhard1954,Bivin}. We then estimate the magnitude of these effects for two representative systems: a weakly coupled electron--proton plasma and an effective $\mathrm{Li}^{+}$--solvated-electron plasma in liquid ammonia. The second example is used as an exploratory condensed-phase realization of the crossover regime; for this first estimate the ammonia background is represented by a constant dielectric permittivity and its frequency dispersion is neglected.

\begin{figure}[ht]
\centering
\includegraphics[width=1.0\textwidth]{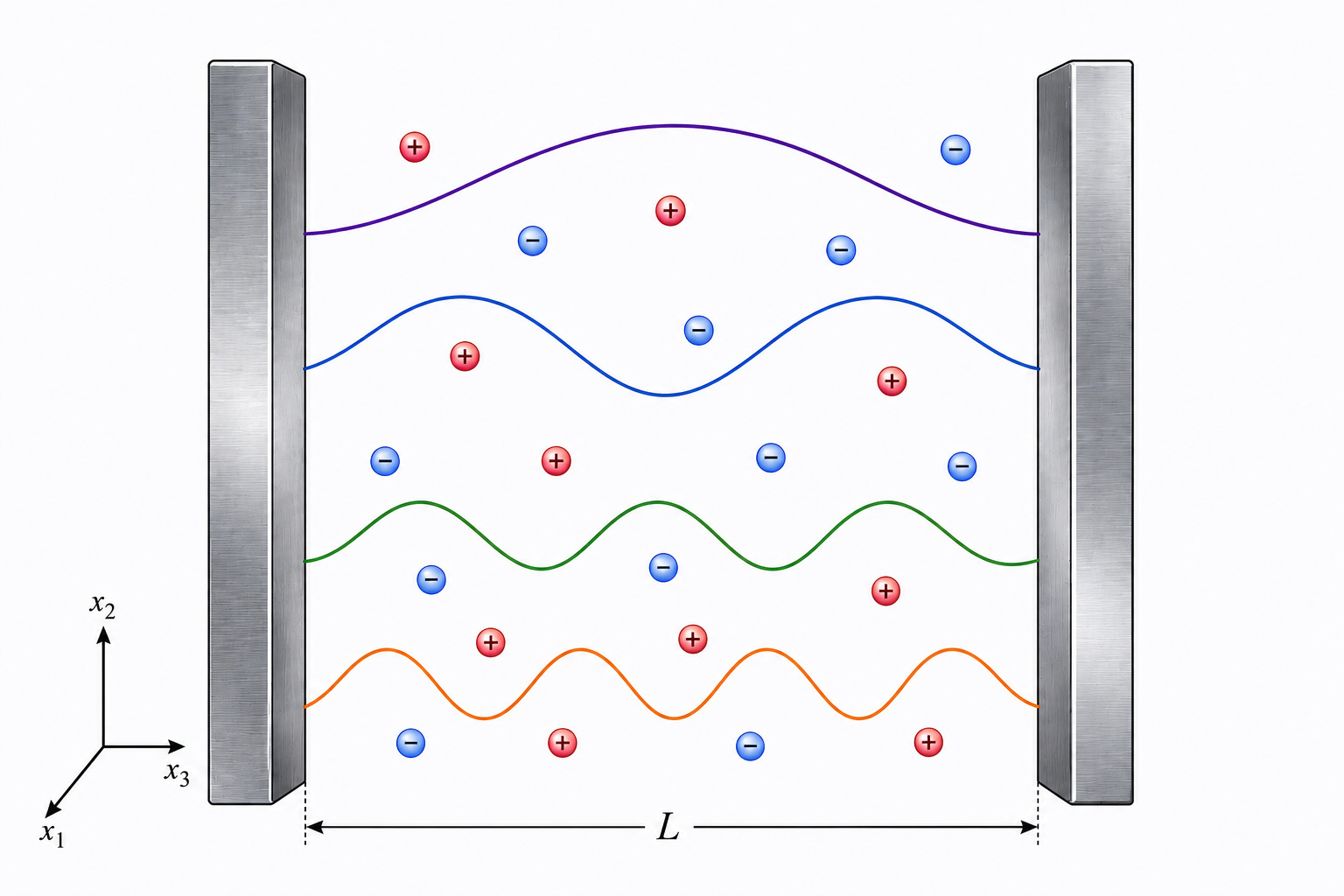}
\caption{Schematic representation of a two-component plasma confined between two parallel perfectly conducting plates separated by a distance $L$.}
\label{Fig1}
\end{figure}

\section{Model}
We consider a two-component Coulomb plasma confined in a slit-like pore of
width $L$. The plasma consists of $N_{+}$ positively charged particles and
$N_{-}$ negatively charged particles with charges
\begin{equation}
q_{+}=q>0,
\qquad
q_{-}=-q,
\end{equation}
respectively. The confining walls are perfectly conducting and electrically
neutral. The particle numbers are fixed and the mobile component is globally
electroneutral,
\begin{equation}
q_{+}N_{+}+q_{-}N_{-}=0.
\end{equation}
For the symmetric charge case considered below this implies $N_{+}=N_{-}$ and,
in the homogeneous bulk region, $n_{+}=n_{-}=n$.

The charged particles are embedded in a homogeneous, isotropic, structureless
dielectric background characterized by a relative permittivity
$\varepsilon_b$. Throughout the present work the background is taken to be
nondispersive. This approximation is useful for isolating the plasma-induced
quantum correction, but it is not intended as a complete material model for a
polar liquid. A frequency-dependent dielectric response can be incorporated by replacing
$\varepsilon_b$ with $\varepsilon_b(i\omega_n)$ in the quadratic electromagnetic
kernel; in the imaginary-time representation this corresponds to a response
that is nonlocal in imaginary time. We leave this extension for future work in
order to isolate the plasma-induced quantum correction. The complete system, including the mobile charges, the conducting
boundaries, the dielectric background, and the quantum electromagnetic field,
is assumed to be in thermal equilibrium at temperature $T$. A schematic
representation is shown in Fig.~\ref{Fig1}.

We describe the electromagnetic field using the finite-temperature
functional-integral formalism, while the quantum motion of the charged
particles is represented by imaginary-time path integrals. The approximations
used below can be organized in terms of three dimensionless parameters.

First, exchange and quantum degeneracy are neglected and the charged particles
are treated, to leading order, by Maxwell--Boltzmann statistics. Here
$m_\alpha$ and $n_\alpha$ denote, respectively, the particle mass and the bulk
number density of species $\alpha$. We use the worldline length
\begin{equation}
\lambda_{dB,\alpha}
=
\frac{\hbar}{\sqrt{m_\alpha k_BT}},
\qquad
\mathcal D_\alpha
\equiv
n_\alpha\lambda_{dB,\alpha}^{3}.
\label{eq:degeneracy_parameter}
\end{equation}
Here $\lambda_{dB,\alpha}$ measures the characteristic extent of the
imaginary-time worldline, while $\mathcal D_\alpha$ is a dimensionless
worldline-overlap parameter. The conventional thermal wavelength is
$\Lambda_\alpha=\sqrt{2\pi}\,\lambda_{dB,\alpha}$. For a species with
internal degeneracy $g_\alpha$, the phase-space degeneracy parameter that
controls exchange corrections is
\begin{equation}
\mathcal D_\alpha^{(\mathrm{ph})}
=\frac{n_\alpha\Lambda_\alpha^3}{g_\alpha}
=\frac{(2\pi)^{3/2}}{g_\alpha}\mathcal D_\alpha.
\end{equation}
The strict Maxwell--Boltzmann regime requires
$\mathcal D_\alpha^{(\mathrm{ph})}\ll1$; for fermions this is equivalent,
up to a numerical factor of order unity, to $E_{F,\alpha}/(k_BT)\ll1$.
We shall also use one parameter set closer to the boundary of this regime as
an exploratory crossover estimate. Importantly, small phase-space degeneracy
does not require the worldline corrections to the electromagnetic response to
vanish. Those corrections are controlled locally by combinations such as
$\lambda_{dB,\alpha}k$ and can therefore become observable before strong
degeneracy sets in. Here and below, $k=|\mathbf{k}|$ denotes the magnitude of
the spatial wave vector of an electromagnetic mode.

Second, we neglect the modification of the particle worldline measure by the
confining boundaries. This approximation requires
\begin{equation}
\frac{\lambda_{dB,\alpha}}{L}\ll1.
\end{equation}
Accordingly, the particle path integrals are evaluated with the free Gaussian
worldline measure. This approximation concerns only the particle trajectories.
The confinement of the electromagnetic field is retained explicitly through
the conducting boundary conditions and the discrete normal spectrum
\begin{equation}
k_3=\frac{\pi l}{L}.
\end{equation}

Third, electrostatic correlations are characterized by the coupling parameter
\begin{equation}
\Gamma
=
\frac{q^2}
{4\pi\varepsilon_0\varepsilon_b\bar r\,k_BT},
\qquad
\bar r=n^{-1/3}.
\label{eq:Gamma}
\end{equation}
The parameters $\Gamma$ and $\mathcal D_\alpha^{(\mathrm{ph})}$ have different physical meanings: $\Gamma$ measures the strength of Coulomb correlations, whereas $\mathcal D_\alpha^{(\mathrm{ph})}$ measures phase-space degeneracy. The Gaussian approximation
to the electromagnetic effective action is quantitatively controlled for
$\Gamma\ll1$. In addition to this weak-coupling regime, we shall use one
case with $\Gamma\simeq0.5$ as an exploratory estimate of the crossover
region. Results in that case should be viewed as semi-quantitative rather than
as a controlled weak-coupling expansion.

For quantitative reference in Sec.~V, we use $\delta=0.1$ to give a concrete
meaning to the symbol $\ll$; this convention is only a small-parameter
benchmark and should not be interpreted as a guaranteed $10\%$ accuracy of an
observable. The corresponding bounds are
\begin{equation}
T\geq T_{\rm MB,\alpha}
=\frac{2\pi\hbar^2}{m_\alpha k_B}
\left(\frac{n_\alpha}{g_\alpha\delta}\right)^{2/3},
\qquad
L\geq L_{\rm b,\alpha}
=\frac{\lambda_{dB,\alpha}}{\delta},
\end{equation}
and, for the weak-coupling condition,
\begin{equation}
T\geq T_\Gamma
=\frac{q^2 n^{1/3}}{4\pi\varepsilon_0\varepsilon_b k_B\delta},
\qquad
n\leq n_\Gamma
=\left(\frac{4\pi\varepsilon_0\varepsilon_b k_BT\delta}{q^2}\right)^3.
\end{equation}
These relations make explicit that the validity conditions constrain the joint
parameter space $(T,n,L)$ rather than defining independent universal minimum
values of all three variables.

The electromagnetic field satisfies the standard perfect-conductor boundary
conditions,
\begin{equation}\label{Boundary1}
\mathbf{E}_{\parallel}(x_1,x_2,0)
=
\mathbf{E}_{\parallel}(x_1,x_2,L)
=
0,
\end{equation}
and
\begin{equation}\label{Boundary2}
\mathbf{B}_{\perp}(x_1,x_2,0)
=
\mathbf{B}_{\perp}(x_1,x_2,L)
=
0.
\end{equation}
The dielectric background changes the electromagnetic propagation inside the
slit but does not modify these boundary conditions.

Finally, the mobile particles are treated as structureless point charges. We
neglect spatially extended intraparticle charge distributions, permanent or
induced multipole moments associated with internal degrees of freedom, and
transitions between internal energy levels. A particularly simple realization
is a nondegenerate electron--proton plasma,
\begin{equation}
q_{+}=e,\qquad q_{-}=-e,
\qquad
m_{+}=m_p,\qquad m_{-}=m_e.
\end{equation}
In this case the theory describes a constrained plasma state at fixed electron and proton
numbers; chemical ionization--recombination equilibrium is not imposed.

\section{Casimir screening}
One introduces the one-particle functional
\begin{equation}
\begin{aligned}
\Theta_\alpha =  \int \mathscr{D}\mu_\alpha \; \bigg(e^{-\frac{1}{\hbar}\int\limits_{0}^{\hbar\beta}d\tau \;\mathscr{U}_\alpha}-1\bigg),
\end{aligned}
\end{equation}
with the normalized path-integration measure
\begin{equation}
\begin{aligned}
\int\mathscr{D}\mu_\alpha (\cdot) = \langle\cdot\rangle_\alpha =  \int \frac{\mathscr{D} \boldsymbol{\xi}}{C^\alpha_0} e^{-\frac{1}{\hbar}\int\limits_{0}^{\hbar\beta}d\tau\;\frac{m_{\alpha}\dot{\boldsymbol{\xi}}^2(\tau)}{2}}(\cdot),
\end{aligned}
\end{equation}
\begin{equation}\label{2}
\begin{aligned}
\mathscr{U}_\alpha=iq_\alpha \bigg(\varphi(\tau,\mathbf{r})-\dot{\mathbf{r}}\cdot\mathbf{A}(\tau,\mathbf{r})\bigg)
\end{aligned}
\end{equation}
Here $A_{\mu}=(-\varphi/c,\mathbf{A})$ is the electromagnetic gauge potential, $C^\alpha_0=\int \mathcal{D}\boldsymbol{\xi}\, e^{-\frac{1}{\hbar}\int\limits_{0}^{\hbar\beta}d\tau\,m_{\alpha}\dot{\boldsymbol{\xi}}^2(\tau)/2}$ is the normalization factor, and $\beta=(k_BT)^{-1}$. The particle coordinate is decomposed as $\mathbf{r}=\mathbf{x}+\boldsymbol{\xi}(\tau)$, where $\mathbf{x}$ is the imaginary-time centroid,
\begin{equation}
\mathbf{x}=\frac{1}{\hbar\beta}\int_0^{\hbar\beta} d\tau \; \mathbf{r}.
\end{equation}
The fluctuation $\boldsymbol{\xi}(\tau)$ describes the quantum spreading about this centroid. 
Then one can consider the expansion
\begin{equation}
\Theta_\alpha
=
\Theta_\alpha^{(1)}+\Theta_\alpha^{(2)}+\mathcal{O}(A^3),
\end{equation}
where the coefficients are
\begin{equation}
\Theta_\alpha^{(1)}
=
-\frac1\hbar
\int_0^{\hbar\beta}d\tau\,
\left\langle\mathscr U_\alpha(\tau)\right\rangle_\alpha,
\end{equation}
\begin{equation}
\Theta_\alpha^{(2)}
=
\frac1{2\hbar^2}
\int_0^{\hbar\beta}d\tau
\int_0^{\hbar\beta}d\tau'\,
\left\langle
\mathscr U_\alpha(\tau)\mathscr U_\alpha(\tau')
\right\rangle_\alpha.
\end{equation}
\par
Then the partition function has the following form~\cite{budkovBrandyshev2025PRE} (see Appendix \ref{Canonical pf})
\begin{equation}\label{1}
\begin{aligned}
 Z=Z_0\int \mathscr{D}A \; \Delta_{F}\; e^{ \frac{S}{\hbar}} ,\quad Z_0=Z_N[0],
\end{aligned}
\end{equation}
and the action is
\begin{equation}\label{5}
S = - \frac{\varepsilon_0}{2} \int\limits_{0}^{\hbar\beta}d\tau\int_V d\mathbf{x}\; \bigg(\varepsilon_b\, \mathbf{E}^2+c^2\,\mathbf{B}^2\bigg) \; + \hbar\;\mathcal W[A] \; + S_{\rm gf},
\end{equation}
where
\begin{equation}
E_{i}=-\partial_i \varphi -\partial_\tau A_i,\quad B_i=\varepsilon_{ijk}\partial_jA_k,
\end{equation}
\begin{equation}
\begin{aligned}
\mathcal W[A]
={}&
\sum_\alpha n_\alpha\int_Vd^3x\,\Theta_\alpha^{(1)}
+
\sum_\alpha n_\alpha\int_Vd^3x\,\Theta_\alpha^{(2)}
\\
&-
\frac12\sum_\alpha\frac{n_\alpha}{V}
\left[
\int_Vd^3x\,\Theta_\alpha^{(1)}
\right]^2
+\mathcal{O}(A^3).
\end{aligned}
\end{equation}

We choose the gauge-fixing term
\begin{equation}
S_{\rm gf}=-\frac{1}{2\mu_0\alpha} \int_{0}^{\hbar\beta}d\tau\int_V d\mathbf{x}\;\bigg(\nabla\cdot\mathbf{A}-\frac{\varepsilon_b}{c^2}\,\partial_\tau\,\varphi\bigg)^2,   
\end{equation}
and use the generalized Feynman gauge, $\alpha=1$, where $\mu_0$ is the vacuum magnetic permeability. Up to a total divergence that vanishes at the boundaries, the action becomes
\begin{equation}
 S = \frac{\varepsilon_0}{2} \int\limits_{0}^{\hbar\beta}d\tau\int_V d\mathbf{x}\; \bigg(\varepsilon_b\, \varphi D \varphi+c^2A_{i}D A_{i}\bigg) \; + \hbar\; \mathcal W[A],
\end{equation}
where we have introduced the operator
\begin{equation}
D=\Delta+\frac{\varepsilon_b}{c^2}\,\partial_\tau{}^2.
\end{equation}
The corresponding Faddeev--Popov determinant is
\begin{equation}
\Delta_{F}=\operatorname{Det}\bigg(-\frac{\delta D \theta}{\delta \theta}\bigg),
\end{equation}
which, up to an overall multiplicative constant, gives
\begin{equation}
\Delta_{F}=\exp\bigg( \sum_{\substack{n,\,l_1,\,l_2\in \mathbb{Z}\\ l\in \mathbb{N}}} \ln( k^{2}c^2 +\varepsilon_b\,\omega_n^2)\bigg),
\end{equation}
\begin{equation}
k^2=k_1^2+k_2^2+k_3^2,\quad k_1 = \frac{2\pi l_{1}}{L_1} , \quad k_2 = \frac{2\pi l_{2}}{L_2}, \quad k_3 = \frac{\pi l}{L}, \quad \omega_n=\frac{2\pi n k_B T}{\hbar},
\end{equation}
where $\mathbb{N}=\{1,2,3,\ldots\}$. The boundary conditions are of Dirichlet type for $A_s$, $s=0,1,2$, and of Neumann type for $A_3$,
\begin{equation}\label{3}
\begin{aligned}
A_{s}(\tau,\mathbf{x})=0, \quad \partial_3A_{3}(\tau,\mathbf{x})=0\quad \text{for}\quad x_3=0,L,\quad s=0,1,2,
\end{aligned}
\end{equation}
These conditions are compatible with the boundary conditions (\ref{Boundary1}) and (\ref{Boundary2}).\par
Performing the functional integral in Eq.~(\ref{1}) in the Gaussian approximation gives
\begin{equation}\label{eq:partition_factorization}
\begin{aligned}
Z=\Delta_F\, Z_M\prod_{j=0}^{4}Z_j,
\end{aligned}
\end{equation}
\begin{equation}\label{PF1}
\begin{aligned}
Z_1=\exp\bigg( -\frac{1}{2}\sum_{\substack{n,\,l_1,\,l_2\in \mathbb{Z}\\ l\in \mathbb{N}}} \ln\bigg( k^{2}c^2 +\varepsilon_b\,\omega_n^2 
+ \bigg(1+\frac{\varepsilon_b\,\omega_n^2}{k^2c^2}\bigg)\chi_1c^2\bigg)\bigg),
\end{aligned}
\end{equation}
\begin{equation}\label{PF2}
\begin{aligned}
Z_2=\exp\bigg( -\sum_{\substack{n,\,l_1,\,l_2\in \mathbb{Z}}} \sideset{}{'}\sum_{l\in \mathbb{N}_0} \ln\bigg( k^{2}c^2 +\varepsilon_b\,\omega_n^2 
+ \chi_2\bigg)\bigg),
\end{aligned}
\end{equation}
\begin{equation}\label{PF3}
\begin{aligned}
Z_3=\exp\bigg( -\frac{1}{2}\sum_{\substack{n,\,l_1,\,l_2\in \mathbb{Z}\\ l\in \mathbb{N}}} \ln\bigg( k^{2}c^2 +\varepsilon_b\,\omega_n^2 \bigg)\bigg).
\end{aligned}
\end{equation}
\begin{equation}
\begin{aligned}
Z_4 = \exp\bigg(-\frac{\beta\mathcal{A}}{2\varepsilon_0\varepsilon_b} \bigg[\frac{8}{L}\sum_{l=0}^{\infty} \frac{\gamma^2(p_{l})}{p_{l}^2(p_{l}^2+\chi_0(p_{l}))}+
\mathbf t^TM^{-1}\mathbf t\bigg]\bigg)
\end{aligned}
\end{equation}
\begin{equation}
Z_{M}
=\exp\bigg(
-\frac12\ln\det M\bigg).
\end{equation}

\begin{equation}
p_{l}=\frac{\pi (2l+1)}{L},\quad \mathcal{A}=L_1L_2,
\end{equation}

\begin{equation}
\gamma(p_{l})=\sum_\alpha n_\alpha q_\alpha e^{-\frac{\lambda_{dB,\alpha}^2p_{l}^2}{24}},
\end{equation}
\begin{equation}
\chi_0(p_{l})=\chi_1(0,p_{l}).
\end{equation}
Here $\mathbb{N}_0=\mathbb{N}\cup\{0\}$, and the primed sum means that the $l=0$ term is included with weight $1/2$. The isolated global $A_3$ zero eigenvalue at $n=l_1=l_2=l=0$ is omitted from the determinant as an overall flat-mode normalization; it does not contribute to the normalized Casimir free energy.

The factors in Eq.~(\ref{eq:partition_factorization}) have the following physical meaning. $Z_1$ is the
longitudinal electric sector, $Z_2$ contains the two transverse
electromagnetic polarizations, and $Z_3$ is the unphysical gauge sector that
must be combined with the Faddeev--Popov determinant. The factor $Z_4$ is the
contribution generated by the quantum-induced linear charge-density source
after Gaussian integration and includes the canonical correction. Finally,
$Z_M$ is a finite-dimensional canonical determinant; as shown below, its
contribution to the free energy per unit area vanishes in the thermodynamic
limit.

The following response functions are introduced:
\begin{equation}
\begin{aligned}
\chi_1(\omega_n,k)=\sum_{\alpha}\kappa^2_{D,\alpha}I_\alpha(\omega_n,k),\quad \kappa^2_{D,\alpha}=\frac{\beta\; q_\alpha^2 n_\alpha}{\varepsilon_0\varepsilon_b},
\end{aligned}
\end{equation}
The total inverse Debye screening length is therefore
\begin{equation}
\kappa_D^2=\sum_\alpha\kappa_{D,\alpha}^2,
\qquad
\lambda_D=\kappa_D^{-1}.
\end{equation}
\begin{equation}
\begin{aligned}
\chi_2(\omega_n,k)=\sum_{\alpha}\omega^2_{p,\alpha}(1-I_\alpha(\omega_n,k)),\quad  \omega^2_{p,\alpha}=\frac{q_\alpha^2 n_\alpha}{\varepsilon_0m_\alpha},
\end{aligned}
\end{equation}
Here $\omega_{p,\alpha}^2$ is the bare plasma parameter used in the response kernel. In a homogeneous dielectric background the collective long-wavelength plasma frequency satisfies $\omega_{\mathrm{pl}}^2=\varepsilon_b^{-1}\sum_\alpha\omega_{p,\alpha}^2$.
\begin{equation}
\begin{aligned}
I_\alpha(\omega_n,k)=2\int_{0}^{\frac{1}{2}} ds \; \cos{\bigg(\hbar\beta\omega_n\bigg(s+\frac{1}{2}\bigg)\bigg)}\;\exp\bigg(\frac{\lambda_{dB,\alpha}^2 k^2}{2}\bigg(s^2-\frac{1}{4}\bigg)\bigg),
\end{aligned}
\end{equation}
Overall multiplicative constants in the Gaussian determinants that are independent of the plate separation have been omitted, since they contribute only an $L$-independent normalization to the free energy and do not affect the Casimir interaction.
\par
The matrix $M$ and the vector $\mathbf t$ are defined by (see Appendix \ref{Source})
\begin{equation}
M_{\alpha\beta}
=
\delta_{\alpha\beta}
-
\frac{8\kappa_{D,\alpha}\kappa_{D,\beta}}{L^2}
\sum_{l=0}^{\infty}
\frac{E_\alpha(p_l)E_\beta(p_l)}{p_l^2(p_{l}^2+\chi_0(p_{l}))},
\end{equation}
\begin{equation}
\begin{aligned}
t_\alpha
=
\frac{8\kappa_{D,\alpha}}{L^{3/2}}
\sum_{l=0}^{\infty}
\frac{E_\alpha(p_l)\gamma(p_l)}{p_l^2(p_{l}^2+\chi_0(p_{l}))},\quad E_\alpha(p)=
\exp\left(-\frac{\lambda_{dB,\alpha}^2p^2}{24}\right).
\end{aligned}
\end{equation}
Then
\begin{equation}
\mathbf t^TM^{-1}\mathbf t
=
\frac{
M_{22}t_1^2-2M_{12}t_1t_2+M_{11}t_2^2
}{
M_{11}M_{22}-M_{12}^2
}.
\end{equation}
In the thermodynamic limit $\mathcal{A}\rightarrow\infty$, the in-plane mode sum becomes
\begin{equation}
\sum_{\substack{l_1,\,l_2\in \mathbb{Z}}} \rightarrow \mathcal{A}\int \frac{d^2 k_{\parallel}}{(2\pi)^2}.
\end{equation}
At fixed $L$, $T$, and species densities, the regularized
fluctuation free energy and the source free energy are
extensive in the plate area. By contrast, $\ln Z_M=-\tfrac12\ln\det M$ is independent of the area. The contribution of $Z_M$ to the free energy density is defined by
\begin{equation}
\frac{F_M}{\mathcal{A}}=-\frac{1}{\beta\mathcal{A}}\ln Z_M \rightarrow 0.
\end{equation}
Consequently, $Z_M$ does not contribute to the free energy
per unit area in the limit $\mathcal A\to\infty$, provided
that $M$ remains nonsingular.

\section{Electromagnetic response of the quantum Boltzmann plasma}
As a consistency check against standard plasma-response formulas, we now consider the vacuum-background limit $\varepsilon_b=1$. For convenience, throughout this section we use units $\hbar=c=k_B=1$, so that the thermal de Broglie length is
\begin{equation}
\begin{aligned}
\lambda_{dB,\alpha}^2=\frac{\beta}{m_\alpha}.
\end{aligned}
\end{equation}
For $\lambda_{dB,\alpha}k\ll1$, expansion of $I_\alpha$ through fourth order in the dimensionless variable $\lambda_{dB,\alpha}k$ gives

\begin{equation}
\begin{aligned}
I_\alpha(\omega_n,k)
=
\frac{\lambda_{dB,\alpha}^{\,2}k^2}
{\beta^2\omega_n^2}
-
3
\frac{\lambda_{dB,\alpha}^{\,4}k^4}
{\beta^4\omega_n^4}
+
\mathcal{O}\!\left((\lambda_{dB,\alpha}k)^6\right),
\qquad n\neq 0.
\end{aligned}
\end{equation}

\begin{equation}
\begin{aligned}
I_\alpha(0,k)
=
1-\frac{\lambda_{dB,\alpha}^2 k^2}{12}
+\frac{\lambda_{dB,\alpha}^4 k^4}{240}
+\mathcal{O}\!\left((\lambda_{dB,\alpha}k)^6\right).
\end{aligned}
\end{equation}
For $n\neq0$ this gives
\begin{equation}
\begin{aligned}
\chi_1(\omega_n,k)
=
\frac{k^2}{\omega_n^2}
\sum_\alpha
\bigg(
\omega^2_{p,\alpha}
\bigg(
1
-
\frac{3\lambda_{dB,\alpha}^2k^2}{\beta^2\omega_n^2}
+\mathcal{O}\!\left((\lambda_{dB,\alpha}k)^4\right)\bigg)\bigg),
\end{aligned}
\end{equation}

\begin{equation}
\begin{aligned}
\chi_2(\omega_n,k)
=
\sum_\alpha\omega_{p,\alpha}^2
\bigg(
1
-
\frac{\lambda_{dB,\alpha}^2k^2}{\beta^2\omega_n^2}
+
\frac{3\lambda_{dB,\alpha}^4k^4}{\beta^4\omega_n^4}
+\mathcal{O}\!\left((\lambda_{dB,\alpha}k)^6\right)
\bigg),
\end{aligned}
\end{equation}
where we used $\omega^2_{p,\alpha}=\kappa^2_{D,\alpha}\lambda_{dB,\alpha}^2/\beta^2$. For $n=0$ one obtains
\begin{equation}
\begin{aligned}
\chi_1(0,k)
=
\sum_\alpha\kappa_{D,\alpha}^2
\bigg(
1
-
\frac{\lambda_{dB,\alpha}^2k^2}{12}
+
\frac{\lambda_{dB,\alpha}^4k^4}{240}
+\mathcal{O}\!\left((\lambda_{dB,\alpha}k)^6\right)
\bigg).
\end{aligned}
\end{equation}

\begin{equation}\label{chi20}
\begin{aligned}
\chi_2(0,k)
=
\sum_\alpha\bigg(
\omega_{p,\alpha}^2
\bigg(
\frac{\lambda_{dB,\alpha}^2k^2}{12}
-
\frac{\lambda_{dB,\alpha}^4k^4}{240}
+
\mathcal{O}\!\left((\lambda_{dB,\alpha}k)^6\right)\bigg)\bigg).
\end{aligned}
\end{equation}

The longitudinal dielectric function may then be introduced through
\begin{equation}
\begin{aligned}
Z_1=\exp\bigg( -\frac{1}{2}\sum_{\substack{n,\,l_1,\,l_2\in \mathbb{Z}\\ l\in \mathbb{N}}} \ln\bigg( \varepsilon_L(\omega_n,k)\bigg(k^{2} +\omega_n^2\bigg) 
\bigg)\bigg),
\end{aligned}
\end{equation}
For $n\neq0$, the leading term in $\lambda_{dB,\alpha}k$ yields the standard plasma form
\begin{equation}
\begin{aligned}
\varepsilon_L(\omega_n,k)
= 1 +
\sum_\alpha\bigg(
\frac{\omega^2_{p,\alpha}}{\omega_n^2}
\bigg(1
+
\mathcal{O}\bigg(\lambda_{dB,\alpha}^2k^2  \bigg)
\bigg)\bigg),
\end{aligned}
\end{equation}
whereas for $n=0$ one recovers the Debye--Huckel form
\begin{equation}
\begin{aligned}
\varepsilon_L(0,k)=1+\sum_{\alpha}
\bigg(
\frac{\kappa^2_{D,\alpha}}{k^2}
\bigg(
1+\mathcal{O}\left(\lambda_{dB,\alpha}^2k^2  \right)
\bigg)\bigg).
\end{aligned}
\end{equation}
Retaining the first nonvanishing quantum correction gives
\begin{equation}
\begin{aligned}
\varepsilon_L(0,k)=1+\sum_{\alpha}\bigg(\frac{\kappa^2_{D,\alpha}}{k^2}\bigg(1-\frac{k^2}{12\kappa_{dB,\alpha}^2}+\mathcal{O}\bigg(\frac{k^4}{\kappa_{dB,\alpha}^4}  \bigg)\bigg)\bigg),
\end{aligned}
\end{equation}
where $\kappa_{dB,\alpha}=1/\lambda_{dB,\alpha}$. This expansion has the same small-$k$ structure as the static Lindhard dielectric function of a degenerate Fermi gas in the random-phase approximation~\cite{Lindhard1954},
\begin{equation}
\begin{aligned}
\varepsilon_L(0,k)=1+\sum_{\alpha}\bigg(\frac{k^2_{TF,\alpha}}{k^2}\bigg(1-\frac{k^2}{12k^2_{F,\alpha}}+\mathcal{O}\left(\frac{k^4}{k^4_{F,\alpha}}\right)\bigg)\bigg),
\end{aligned}
\end{equation}
where $k_{TF,\alpha}$ is the Thomas--Fermi wave number and $k_{F,\alpha}$ is the Fermi wave number of species $\alpha$. Thus, to second order in $k$, the two expansions have an analogous structure under the formal correspondence
\begin{equation}
\begin{aligned}
k_{TF,\alpha}\leftrightarrow \kappa_{D,\alpha},\quad k_{F,\alpha} \leftrightarrow \kappa_{dB,\alpha}.
\end{aligned}
\end{equation}
Within the Gaussian theory, $Z_1$ represents one longitudinal electric degree of freedom,
\begin{equation}
\mathbf{E}_L=-i(k\varphi+\omega_nA_L)\hat{\mathbf{k}},\qquad \mathbf{A}_L=A_L\hat{\mathbf{k}},
\end{equation}
where $\mathbf{A}_L\parallel\mathbf{k}$ and $\hat{\mathbf{k}}=\mathbf{k}/k$. The factor $Z_2$ represents two transverse electromagnetic degrees of freedom,
\begin{equation}
\mathbf{E}_T=-i\omega_n\,\mathbf{A}_T,\qquad \mathbf{B}=i\mathbf{k}\times\mathbf{A}_T,
\end{equation}
where $\mathbf{A}_T\perp\mathbf{k}$, whereas $Z_3$ represents one unphysical gauge degree of freedom.
In the same units, the static magnetic permeability is
\begin{equation}
\mu=\lim_{k\to 0} \frac{\mu_0}{1+\frac{\Pi_T(0,k)}{k^2}}.
\end{equation}
Here $\Pi_T$ is the transverse eigenvalue of the polarization operator defined by
\begin{equation}
\begin{aligned}
Z_2=\exp\bigg( -\sideset{}{'}\sum_{\substack{n,\,l_1,\,l_2\in \mathbb{Z}\\ l\in \mathbb{N}_0}} \ln\bigg( k^{2} + \omega_n^2 
+ \Pi_T(\omega_n,k)\bigg)\bigg),
\end{aligned}
\end{equation}
so that $\chi_2=\Pi_T$. For a weak magnetic response, $\lambda_{dB,\alpha}\omega_{p,\alpha}\ll1$, Eq.~(\ref{chi20}) gives
\begin{equation}
\begin{aligned}
\mu
=\mu_0\bigg(1+\chi_{orb}\bigg) + \mu_0\mathcal{O}\!\left((\lambda_{dB,\alpha}\omega_{p,\alpha})^4\right),
\end{aligned}
\end{equation}
where the orbital magnetic susceptibility $\chi_{orb}$ has the form
\begin{equation}
\begin{aligned}
\chi_{orb}
=
-\sum_\alpha
\frac{\omega_{p,\alpha}^2\lambda_{dB,\alpha}^2}{12},
\end{aligned}
\end{equation}
which coincides with the leading high-temperature Landau orbital diamagnetic susceptibility; see Eq.~(1.1b) of Bivin and McClure~\cite{Bivin}.

\section{Numerical results and discussion}

The numerical calculations are intended to estimate the scale of the quantum
correction and to identify a regime in which it becomes visible before the
onset of strong degeneracy. We therefore consider two complementary examples.
The first is a dilute electron--proton plasma, for which both the weak-coupling
and nondegeneracy conditions are satisfied by a wide margin. It provides a
controlled reference case dominated by classical Debye screening. The second
is a dilute lithium solution in liquid ammonia, described at the present level
as an effective two-component plasma of $\mathrm{Li}^{+}$ ions and solvated
electrons. Metal--ammonia solutions provide a well-studied class of liquids containing
solvated electrons and solvated metal cations, with a continuous evolution
from dilute electrolytic behavior toward metallic states as the metal
concentration increases~\cite{cohenThompson1968,zurekEdwardsHoffmann2009}. Ab initio molecular-dynamics studies support the formation of solvated electrons when lithium is dissolved in liquid ammonia~\cite{chabanPrezhdo2016}. This system therefore provides a useful, although necessarily simplified, condensed-phase example in which the electronic quantum length scale is appreciable while the electron component remains approximately nondegenerate.

Numerically, the one-dimensional integrals entering the response functions and
pressure formulas are evaluated by Gauss--Legendre quadrature. For the static
form factor we use its equivalent Dawson-function representation,
\begin{equation}
I_\alpha(0,k)
=\frac{2\sqrt{2}}{\lambda_{dB,\alpha}k}
F_D\!\left(\frac{\lambda_{dB,\alpha}k}{2\sqrt{2}}\right),
\qquad
F_D(x)=e^{-x^2}\int_0^x e^{t^2}\,dt,
\end{equation}
which is convenient for stable evaluation over the relevant wave-vector range.
Convergence is checked by increasing the quadrature order and the Matsubara and
normal-mode cutoffs until the force gradient is unchanged on the scale shown in
the figures. Details of the worldline integration, the canonical source term,
and the pressure representation used in the calculations are collected in
Appendices A, C, and E, respectively.

Throughout this section we use the standard notation of surface-force theory~\cite{israelachvili2011,derjaguinChuraevMuller1987}. For two parallel plates at separation $L$, $\Pi(L)$ denotes the disjoining pressure,
\begin{equation}
\Pi(L)=-\frac{dW(L)}{dL},
\end{equation}
where $W(L)$ is the interaction free energy per unit area and $W(\infty)=0$. In the Derjaguin approximation, valid when the surface separation is much smaller than the radii of curvature, the interaction force between two spheres is
\begin{equation}
F(L)=2\pi R_{\mathrm{eff}}W(L)
=2\pi R_{\mathrm{eff}}\int_L^\infty dh\,\Pi(h),
\qquad
R_{\mathrm{eff}}=\frac{R_1R_2}{R_1+R_2}.
\end{equation}
The sphere--plane limit is obtained by setting $R_{\mathrm{eff}}=R$. Consequently, the force gradient used below is related directly to the planar disjoining pressure by
\begin{equation}
F'(L)=-2\pi R_{\mathrm{eff}}\Pi(L),
\end{equation}
and hence by $F'(L)=-2\pi R\Pi(L)$ for the sphere--plane geometry considered in the numerical examples.

\subsection{Electron--proton plasma}

For the electron--proton plasma we use $T=1200~\mathrm{K}$,
$\varepsilon_b=1$, and equal densities
$n_e=n_p=7.1\times10^{17}~\mathrm{m}^{-3}$. The plate separation is varied
between $4$ and $6~\mu\mathrm{m}$ and the sphere radius used in the
proximity-force conversion is $R=5~\mathrm{mm}$. For these parameters,
\begin{equation}
\Gamma\simeq1.24\times10^{-2},
\qquad
\mathcal D_e=n_e\lambda_{dB,e}^3\simeq4.49\times10^{-10}.
\end{equation}
For spin-$1/2$ electrons this corresponds to the conventional phase-space parameter $\mathcal D_e^{(\mathrm{ph})}\simeq3.54\times10^{-9}$.
The boundary parameter is
\begin{equation}
1.43\times10^{-4}
\lesssim
\frac{\lambda_{dB,e}}{L}
\lesssim
2.15\times10^{-4}.
\end{equation}
The combined Debye length is approximately
$\lambda_D=2.01~\mu\mathrm{m}$, so the calculation probes separations of
order a few Debye lengths. Using the reference threshold $\delta=0.1$
introduced in Sec.~II, the nondegeneracy condition at this density requires
only $T\geq0.0129~\mathrm{K}$, whereas $\Gamma\leq0.1$ requires
$T\geq149~\mathrm{K}$. At $T=1200~\mathrm{K}$ the weak-coupling criterion
allows densities up to $n\simeq3.72\times10^{20}~\mathrm{m}^{-3}$, and the
boundary criterion requires only $L\geq8.58~\mathrm{nm}$. Thus the parameters
used here lie well inside the controlled regime according to all three
criteria.

\begin{figure}[ht]
\centering
\includegraphics[width=0.8\textwidth]{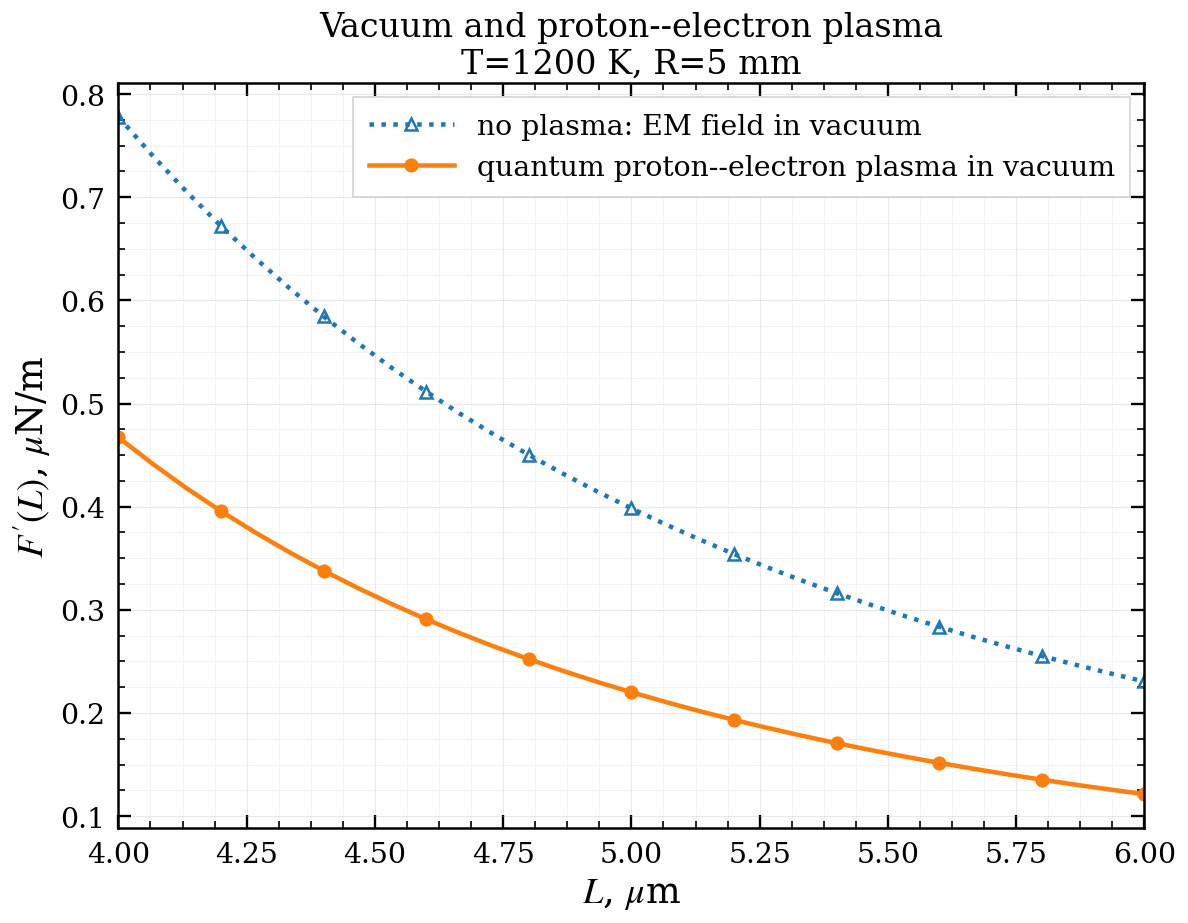}
\caption{Force gradient for a dilute electron--proton plasma at $T=1200~\mathrm{K}$ compared with the electromagnetic result at the same temperature in the absence of mobile charges. The sphere radius is $R=5~\mathrm{mm}$.}
\label{Fig2}
\end{figure}

Figure~\ref{Fig2} compares the total force gradient in the plasma with the
corresponding electromagnetic result at the same temperature in the absence of
mobile charges. The plasma reduces the force gradient by about $40\%$ at the
lower end of the interval and by nearly $50\%$ at the upper end. This strong
suppression is predominantly the classical screening of the static
longitudinal sector. Because both $\mathcal D_e$ and
$\lambda_{dB,e}/L$ are extremely small, the explicit quantum correction in
this example is correspondingly weak. The electron--proton system therefore
serves mainly as a controlled reference demonstrating that the theory connects
smoothly to the classical screening regime.

For the electron--proton plasma, the classical and quantum predictions are
practically indistinguishable on the scale of Fig.~\ref{Fig2}. Over the 
considered separation range, the maximum absolute
difference between the corresponding force gradients is
approximately $0.74~\mathrm{fN/m}$. Thus, although the
formalism fully retains the finite imaginary-time extension of the particle
worldlines, quantum corrections to the Casimir force are essentially
negligible in this dilute high-temperature regime. This provides a useful
reference classical limit for comparison with the lithium--ammonia system,
where the quantum--classical difference becomes appreciable.

\subsection{Lithium in liquid ammonia}

As a second example we consider a dilute lithium solution in liquid ammonia. The lithium--ammonia system therefore represents a qualitatively different
regime: the plasma remains approximately nondegenerate, while quantum delocalization is already sufficiently strong to produce a measurable correction to the force gradient. In the dilute regime, alkali-metal--ammonia solutions are commonly described in terms of solvated metal cations and solvated electrons~\cite{cohenThompson1968,zurekEdwardsHoffmann2009,chabanPrezhdo2016}. Optical measurements on lithium--ammonia solutions also show a pronounced electronic response associated with the solvated-electron sector~\cite{somoanoThompson1970}. Within our minimal model we retain only the two mobile charged species and represent the ammonia solvent by a homogeneous dielectric background.

We choose $T=240~\mathrm{K}$ and $\varepsilon_b=22$, close to the static
permittivity of liquid ammonia in this temperature range~\cite{perkynsKusalikPatey1986}. The dielectric response of real liquid ammonia is, however, frequency dependent: dielectric-relaxation and far-infrared measurements show substantial dispersion of the complex permittivity~\cite{fishMillerSmyth1958,breitschwerdtSchmidt1970,kindtSchmuttenmaer1996}. In the present first estimate we deliberately neglect this dispersion and use the same constant $\varepsilon_b$ for all Matsubara frequencies. Within the present Gaussian formulation, dispersion can be incorporated by replacing $\varepsilon_b$ with $\varepsilon_b(i\omega_n)$ in the electromagnetic kernel; no conceptual modification of the approach is required. We retain a constant background here to isolate the plasma-induced quantum correction, but this simplification prevents a quantitatively precise comparison with a real lithium--ammonia experiment, especially in the nonzero-frequency sector.

The mobile charges are $q_e=-e$ and $q_{\mathrm{Li}^{+}}=+e$, with
$m_{\mathrm{Li}^{+}}=1.16494\times10^{-26}~\mathrm{kg}$. For this exploratory estimate we use the bare electron mass for the solvated-electron kinetic term; solvent-induced mass renormalization and the finite internal structure of the solvated electron are not included. We use equal number densities
\begin{equation}
n_e=n_{\mathrm{Li}^{+}}
=3.943438\times10^{24}~\mathrm{m}^{-3},
\end{equation}
and separations $580~\mathrm{nm}\leq L\leq720~\mathrm{nm}$. The sphere
radius is again $R=5~\mathrm{mm}$. For these parameters,
\begin{equation}
\lambda_{dB,e}=1.919~\mathrm{nm},
\qquad
\lambda_{dB,\mathrm{Li}^{+}}=0.01697~\mathrm{nm},
\end{equation}
while the worldline-overlap parameters in our convention are
\begin{equation}
\mathcal D_e=2.79\times10^{-2},
\qquad
\mathcal D_{\mathrm{Li}^{+}}=1.93\times10^{-8}.
\end{equation}
For spin-$1/2$ electrons, the conventional phase-space degeneracy is
\begin{equation}
\mathcal D_e^{(\mathrm{ph})}
=\frac{(2\pi)^{3/2}}{2}\mathcal D_e
\simeq0.22,
\qquad
\frac{E_F}{k_BT}\simeq0.44.
\end{equation}
Thus the electron component remains on the nondegenerate side of the crossover,
but it is no longer deep in the asymptotic Maxwell--Boltzmann regime. Its
quantum delocalization is also appreciable on microscopic length scales. At the same time,
\begin{equation}
2.67\times10^{-3}
\lesssim
\frac{\lambda_{dB,e}}{L}
\lesssim
3.31\times10^{-3},
\end{equation}
so boundary corrections to the worldline measure remain small. The coupling
parameter and Debye-sphere population are
\begin{equation}
\Gamma=0.50,
\qquad
N_D=0.188.
\end{equation}
For the same reference threshold $\delta=0.1$, the phase-space
nondegeneracy criterion at the chosen density would require
$T\geq406~\mathrm{K}$, while the weak-coupling criterion would require
$T\geq1200~\mathrm{K}$. Equivalently, at $T=240~\mathrm{K}$ the condition
$\Gamma\leq0.1$ would restrict the density to
$n\leq3.15\times10^{22}~\mathrm{m}^{-3}$. The boundary criterion is much less
restrictive, requiring only $L\geq19.2~\mathrm{nm}$. Thus the chosen
lithium--ammonia parameters intentionally lie outside the strict $0.1$
benchmarks for degeneracy and Coulomb coupling, while remaining safely within
the free-worldline boundary criterion.

This is already an intermediate-coupling regime. Moreover, with $\mathcal D_e^{(\mathrm{ph})}\simeq0.22$, Maxwell--Boltzmann statistics should be regarded as a leading-order description: exchange corrections are not parametrically small on the scale of the sub-percent force difference and would have to be retained in a precision calculation. The lithium--ammonia results should therefore be interpreted as an estimate of the magnitude and sign of the quantum correction within the present Gaussian two-component model, not as a quantitatively controlled prediction for the real solution.

\begin{figure}[ht]
\centering
\includegraphics[width=0.8\textwidth]{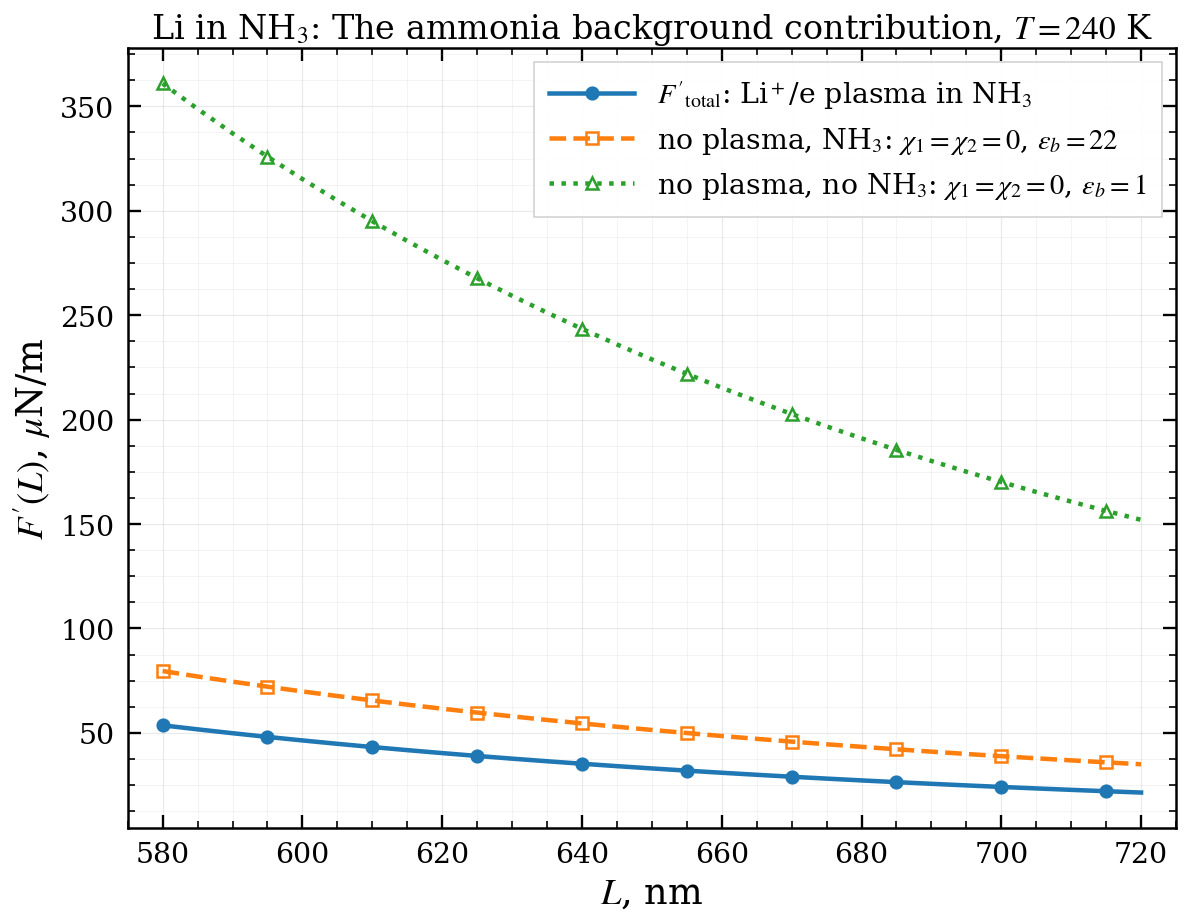}
\caption{Force gradient for the effective $\mathrm{Li}^{+}$--electron plasma in liquid ammonia at $T=240~\mathrm{K}$. The three curves separate the effects of the dielectric background and the mobile charges. The sphere radius is $R=5~\mathrm{mm}$.}
\label{Fig3}
\end{figure}

\begin{figure}[ht]
\centering
\includegraphics[width=0.8\textwidth]{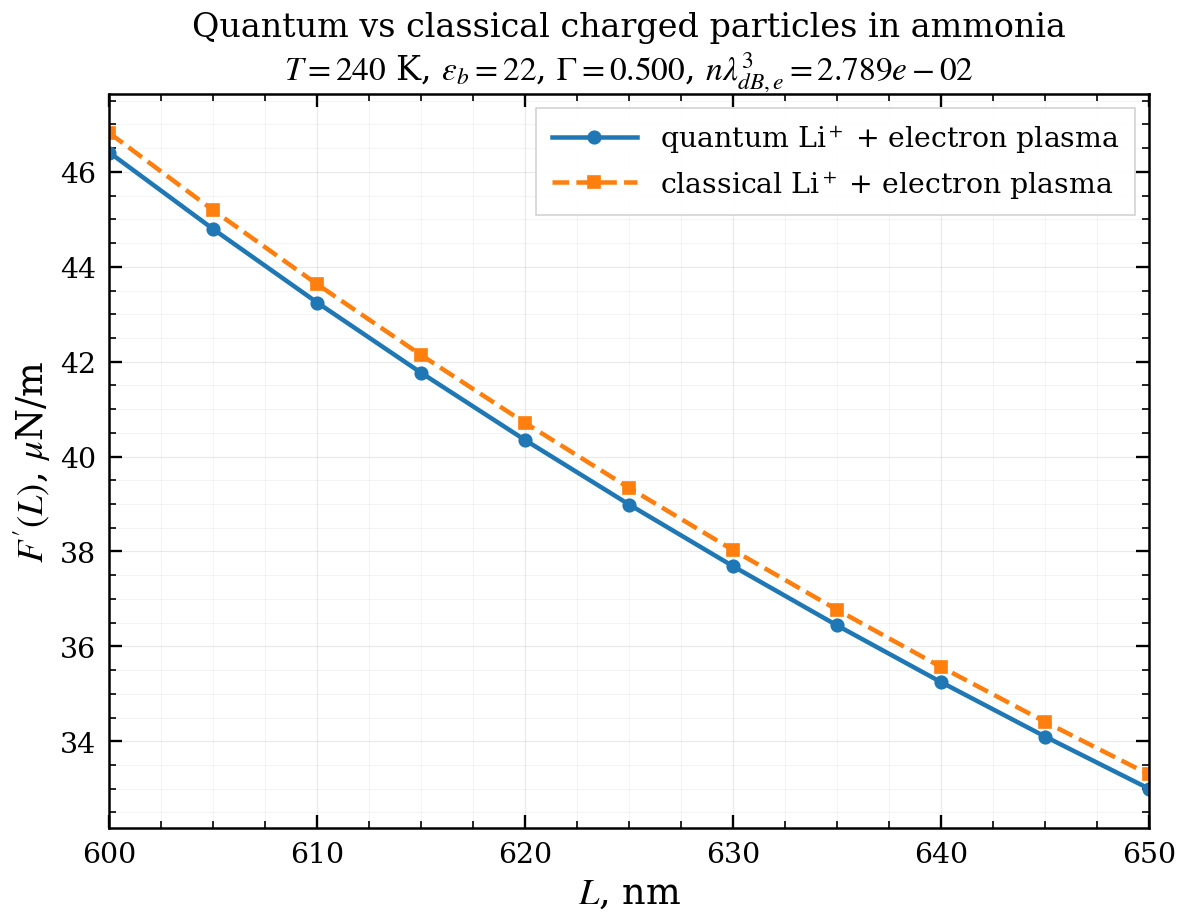}
\caption{Comparison of the quantum and classical $\mathrm{Li}^{+}$--electron plasma predictions in the ammonia background. The annotation $n_e\lambda_{dB,e}^3$ uses the worldline-length convention of Eq.~(\ref{eq:degeneracy_parameter}); the corresponding conventional phase-space degeneracy parameter is $\mathcal D_e^{(\mathrm{ph})}\simeq0.22$.}
\label{Fig4}
\end{figure}

\begin{figure}[ht]
\centering
\includegraphics[width=0.8\textwidth]{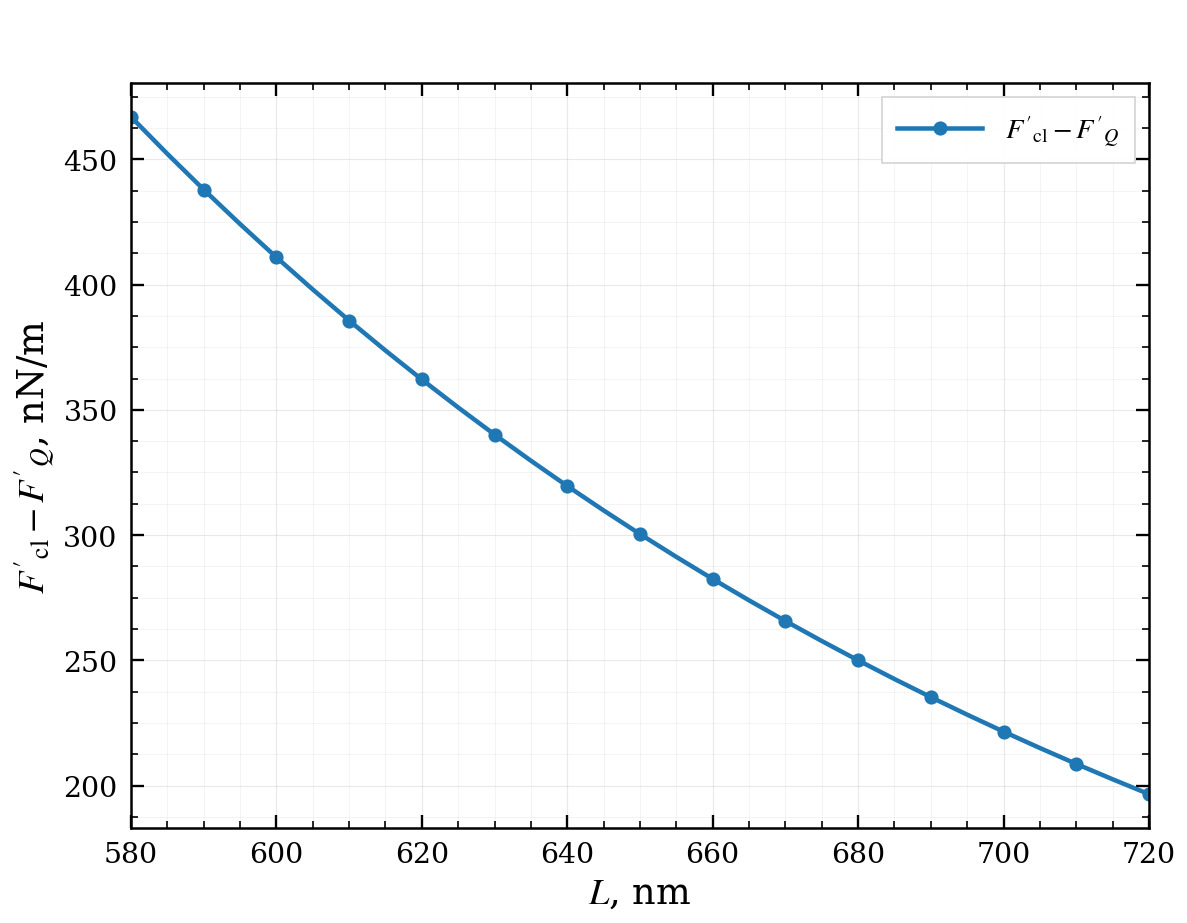}
\caption{Difference between the classical and quantum plasma force gradients for the lithium--ammonia case.}
\label{Fig5}
\end{figure}

Figure~\ref{Fig3} separates the effect of the dielectric background from the
additional effect of the mobile charges. The constant ammonia permittivity
strongly suppresses the force relative to the vacuum-background result, and
the $\mathrm{Li}^{+}$--electron plasma produces a further reduction. The
comparison with a classical plasma is shown on an enlarged scale in
Fig.~\ref{Fig4}. Quantum delocalization decreases the force gradient slightly
relative to the classical result. The difference, shown directly in
Fig.~\ref{Fig5}, varies from approximately $0.467~\mu\mathrm{N/m}$ at
$L=580~\mathrm{nm}$ to $0.197~\mu\mathrm{N/m}$ at
$L=720~\mathrm{nm}$ and remains below one percent of the total force
gradient. Thus these parameters realize the intended crossover regime within
the present model: the plasma remains approximately nondegenerate, while the
finite worldline extension gives a resolvable correction to the Casimir
interaction. The magnitude of this correction is also encouraging from an
experimental point of view. State-of-the-art dynamic-AFM measurements of the
Casimir force gradient between metallic surfaces resolve force-gradient
differences on the scale of $10^{-1}~\mu\mathrm{N/m}$ at separations of
several hundred nanometers~\cite{liu2019precision}. The predicted difference
of approximately $0.197$--$0.467~\mu\mathrm{N/m}$ therefore lies within the
sensitivity range of present force-gradient techniques. This comparison is
intended only as an indication of instrumental sensitivity; a dedicated
measurement in a liquid-ammonia environment would require separate control of
dielectric dispersion, electrostatic backgrounds, and hydrodynamic effects.\par

\begin{figure}[ht]
\centering
\includegraphics[width=0.8\textwidth]{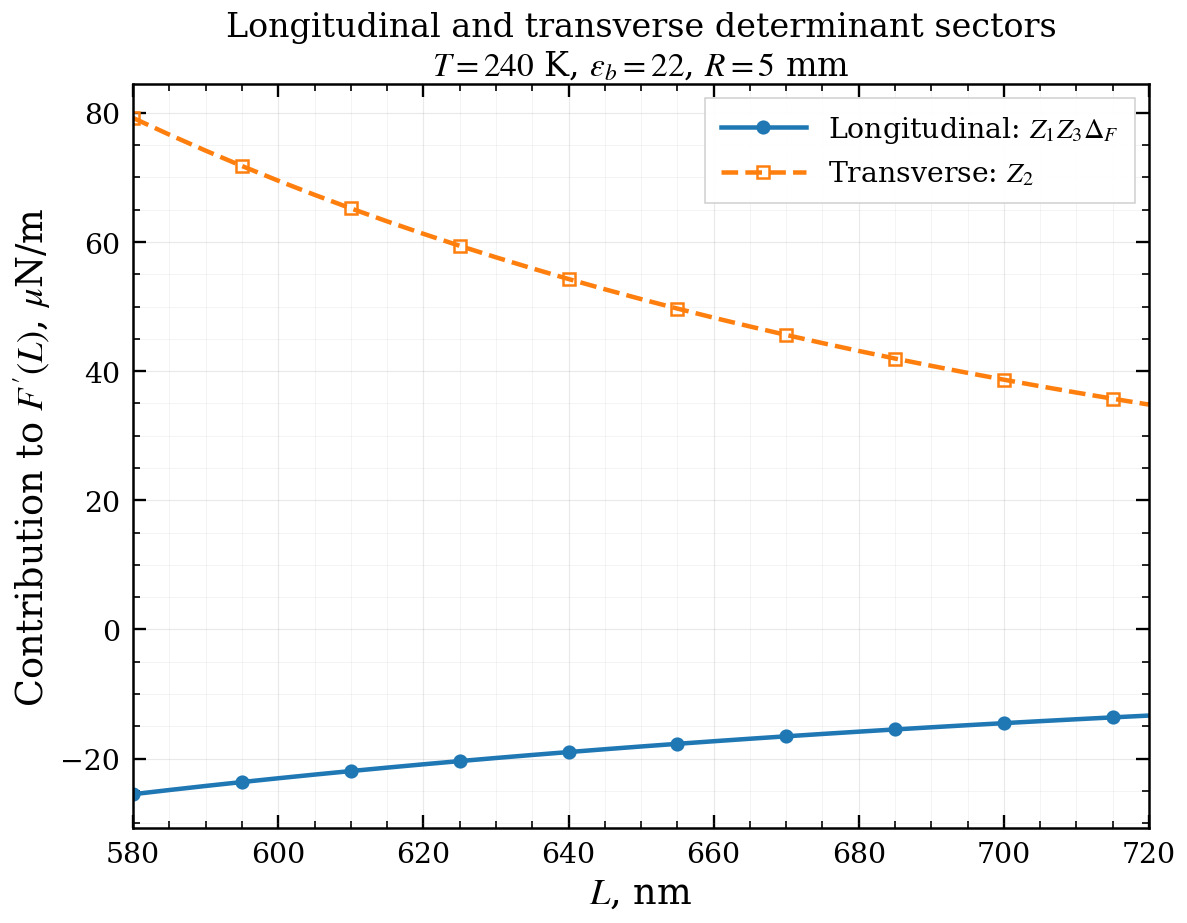}
\caption{Separate longitudinal and transverse contributions
to the Casimir force gradient for the lithium--ammonia case,
after cancellation of the gauge and ghost factors. The source contribution $Z_4$ is not included.}
\label{Fig6}
\end{figure}

Figure~\ref{Fig6} shows the separate longitudinal and transverse electromagnetic
contributions to the Casimir force gradient for lithium in liquid ammonia after the gauge
and ghost factors have been combined and canceled. The transverse contribution is positive,
whereas the longitudinal contribution is negative and therefore reduces the total force
gradient. The transverse contribution dominates throughout the range considered, so the sum
of these two sectors remains positive.

\subsection{General trends}

The two examples clarify the hierarchy of effects. The dominant plasma effect
is the screening of the low-frequency longitudinal contribution. Genuine
quantum corrections enter through both the longitudinal and transverse response
functions $\chi_1$ and $\chi_2$. For the lithium--ammonia parameters,
about $87$--$92\%$ of the observable quantum--classical difference arises from the
transverse response of the nonzero Matsubara modes, while the source contribution $Z_4$
accounts for approximately $8$--$13\%$. For example, at $L=720~\mathrm{nm}$, the difference between the classical
and quantum force gradients is approximately $0.916\%$ of
the total quantum force gradient. The source contribution
$Z_4$ accounts for approximately $0.119\%$ of the total
quantum force gradient, or about $13\%$ of this difference.

The comparison also shows that quantum plasma effects need not require a
degenerate electron gas. They can become visible already for a small but
finite degeneracy parameter $\mathcal D_e$, provided that the relevant
response functions probe wave-vector and frequency scales on which the finite
electron worldline extension matters. This intermediate regime is the intended domain of the present quantum Boltzmann description as a leading-order treatment rather than a precision expansion in the degeneracy parameter. For a realistic
lithium--ammonia system, however, dielectric dispersion and correlation effects
beyond the Gaussian approximation must be included before quantitative
comparison with experiment can be attempted.

\section{Conclusions}

We have developed a finite-temperature field-theoretical description of the
Casimir interaction in a two-component quantum Boltzmann plasma confined
between perfectly conducting boundaries. Treating the charged particles as
imaginary-time worldlines and retaining the electromagnetic effective action
to Gaussian order leads to a transparent decomposition into longitudinal,
transverse, gauge, and quantum-induced source sectors. In the classical limit
the result reduces to the previously established Casimir-screening picture:
the static longitudinal mode is Debye screened, whereas the transverse
electromagnetic fluctuations remain separated from the classical mobile
charges. Finite de Broglie wavelengths break this separation by generating
current fluctuations and modify the nonzero Matsubara modes through the
quantum response functions $\chi_1$ and $\chi_2$.

The response functions also establish two direct connections with standard
quantum-plasma results. In the static long-wavelength expansion, and through
the leading quantum correction, the longitudinal dielectric function has the
same structure as the static Lindhard result under the formal correspondence
\begin{equation}
 k_{\rm TF}\leftrightarrow\kappa_D,
 \qquad
 k_F\leftrightarrow\lambda_{dB}^{-1}.
\end{equation}
Thus, at this order, the Debye wave number replaces the Thomas--Fermi wave
number and the inverse thermal de Broglie length replaces the Fermi wave
vector. The weak static transverse response, in turn, reproduces the leading
high-temperature Landau orbital diamagnetic susceptibility. These two limits
provide independent analytical checks of the worldline response functions and
a systematic route to the leading quantum corrections to the dielectric and
magnetic response of a nondegenerate plasma. In addition, the finite worldline
extension produces the source factor $Z_4$. This term is absent in 
classical theory, but for the geometries considered here, it accounts for 
approximately $8$--$13\%$ of the quantum--classical difference for
the lithium--ammonia case.

The numerical examples illustrate two sides of the classical--quantum
crossover. A dilute electron--proton plasma lies deeply in the weak-coupling,
nondegenerate regime and is dominated by classical screening. In the effective
lithium--ammonia example the phase-space degeneracy parameter is still below
unity, while the quantum correction to the force gradient becomes visible at
the sub-percent level. This demonstrates that quantum-plasma corrections can
arise on the nondegenerate side of the crossover, although exchange effects
must be included before a precision comparison can be attempted. At the same time, the
lithium--ammonia calculation has $\Gamma\simeq0.5$ and is therefore only
semi-quantitative within the Gaussian approximation. The predicted force-gradient difference is nevertheless on the $10^{-1}~\mu\mathrm{N/m}$ scale resolved in modern dynamic-AFM Casimir measurements, which motivates a more realistic material-specific analysis.

For the present first estimate, liquid ammonia was represented by a constant
relative permittivity. Real ammonia has a strongly frequency-dependent
dielectric response, and incorporation of this dispersion, together with
correlations beyond Gaussian order and eventually exchange effects, is a
necessary next step toward a material-specific theory. A quantitative
comparison with experiment will also require replacing the ideal-conductor
boundary conditions by the frequency-dependent response of the actual plate
materials. These extensions should allow one to determine more precisely the
parameter window in which the classical Casimir-screening regime crosses over
to a genuinely quantum plasma response.

\section*{Data availability}
No new data were created or analyzed in this study.

\begin{acknowledgments}
This work was supported by the Russian Science Foundation (Grant No. 24-11-00096). Numerical calculations were made using computational resources of HPC facilities at NRU HSE. We dedicate this work to the memory of Rudi Podgornik, whose ideas and scientific vision inspired us to pursue this problem.
\end{acknowledgments}

\appendix
\section{Calculation of the functional integral}
Throughout this Appendix we use units $\hbar=c=k_B=1$. 
The worldline displacement $\boldsymbol{\xi}$ is expanded as
\begin{equation}
\begin{aligned}
\xi_i(\tau) = \sum_{\substack{n\in \mathbb{N} \\ a=1,2}}a_{ia}(n)u_{a}(\tau,\beta,\omega_n),\quad i=1,2,3
\end{aligned}
\end{equation}
in the orthonormal basis
 \begin{equation}
\begin{aligned}
u_{1}(\tau,\beta,0) = \frac{1}{\sqrt{\beta}},\quad u_{1}(\tau,\beta,\omega_n) = \sqrt{\frac{2}{\beta}}\cos (\omega_n\tau),\quad n\in \mathbb{N},\\
u_{2}(\tau,\beta,\omega_n)=\sqrt{\frac{2}{\beta}}\sin (\omega_n\tau), \quad n\in \mathbb{N}.
\end{aligned}
\end{equation}
The particle coordinate is then written as
\begin{equation}
\begin{aligned}
 r_i(\tau) = \sum_{(n,a)\in K}a_{ia}(n)u_{a}(\tau,\beta,\omega_n),\quad a_{i1}(0)=\sqrt{\beta}x_i,\quad i=1,2,3
 \end{aligned}
\end{equation}
where $K=\{ (n,a)\}\setminus \{ (0,2)\}$, $a=1,2$, $n\in \mathbb{N}_0$.
The gauge potential is expanded in the same temporal basis,
\begin{equation}\label{14}
\begin{aligned}
A_{3}(\tau,\mathbf{x})=\frac{1}{\sqrt{L_1L_2L}}\sum_{\substack{(n,a)\in K\\ l_1,l_2,l \in \mathbb{Z}}}A_{3 a}(n,l_1,l_2,l)u_{a}(\tau,\beta,\omega_n)e^{i\mathbf{k} \cdot \mathbf{x}}.
\end{aligned}
\end{equation}
\begin{equation}
\begin{aligned}
A_{s}(\tau,\mathbf{x})=\frac{1}{\sqrt{L_1L_2L}}\sum_{\substack{(n,a)\in K\\ l_1,l_2,l \in \mathbb{Z}}}A_{s a}(n,l_1,l_2,l)u_{a}(\tau,\beta,\omega_n)e^{i\mathbf{k} \cdot \mathbf{x}}\quad s=0,1,2.
\end{aligned}
\end{equation}
where $\mathbf{k}=(k_1,k_2,k_3)$ and $A_0=-\varphi$. The boundary conditions are encoded in the parity of the normal modes. From Eq.~(\ref{3}) it follows that
\begin{equation}
\begin{aligned}
A_{s a}(n,l_1,l_2,-l)=-A_{s a}(n,l_1,l_2,l),\quad l\in \mathbb{N}_0, \quad s=0,1,2,
\end{aligned}
\end{equation}
\begin{equation}
\begin{aligned}
A_{3 a}(n,l_1,l_2,-l)=A_{3 a}(n,l_1,l_2,l),\quad l\in \mathbb{N}_0.
\end{aligned}
\end{equation}
Thus, the expansion of $A_s$ can be rewritten as
\begin{equation}
\begin{aligned}
A_{s}(\tau,\mathbf{x})=\frac{1}{\sqrt{L_1L_2L}}\sum_{\substack{(n,a)\in K\\ l_1,l_2 \in \mathbb{Z},l \in \mathbb{Z}_0}}A_{s a}(n,l_1,l_2,l)u_{a}(\tau,\beta,\omega_n)e^{i\mathbf{k} \cdot \mathbf{x}},\quad s=0,1,2.
\end{aligned}
\end{equation}
where $\mathbb{Z}_0=\mathbb{Z}\setminus\{0\}$. Since the gauge field is real,
\begin{equation}
A_\mu(\tau,\mathbf{x})=A_\mu^*(\tau,\mathbf{x}),
\qquad \mu=0,1,2,3,
\end{equation}
the spatial Fourier coefficients are not all independent. Combining the reality condition with the odd/even parity in $l$ gives, for example,
\begin{equation}
A_{sa}(n,-l_1,-l_2,l)=-A_{sa}^*(n,l_1,l_2,l),
\qquad s=0,1,2,
\end{equation}
\begin{equation}
A_{sa}(n,-l_1,l_2,l)=-A_{sa}^*(n,l_1,-l_2,l),
\qquad s=0,1,2,
\end{equation}
and
\begin{equation}
A_{3a}(n,-l_1,-l_2,l)=A_{3a}^*(n,l_1,l_2,l),
\end{equation}
\begin{equation}
A_{3a}(n,-l_1,l_2,l)=A_{3a}^*(n,l_1,-l_2,l).
\end{equation}
For generic $l_1,l_2>0$, only two complex coefficients (equivalently, four real coefficients) among the four in-plane sign choices are independent; the corresponding reduced relations apply when one of the in-plane indices vanishes. The functional integral is therefore taken only over independent Fourier amplitudes, with the remaining coefficients eliminated by these constraints.
The normalized Gaussian functional measure over $\boldsymbol{\xi}$ may be written in terms of the Fourier amplitudes as
\begin{equation}
\begin{aligned}
\int\mathscr{D}\mu_\alpha\,F[\boldsymbol{\xi}]
={}&
\prod_{i=1}^{3}\prod_{n=1}^{\infty}
\left[
\frac{m_\alpha\omega_n^2}{2\pi}
\int da_{i1}(n)\,da_{i2}(n)\,
\exp\left\{-\frac{m_\alpha\omega_n^2}{2}
\left[a_{i1}^2(n)+a_{i2}^2(n)\right]\right\}
\right]
\\
&\times
F\left[
\left\{
\sum_{n\geq1}\sum_{a=1}^{2}
a_{ia}(n)u_a(\tau,\beta,\omega_n)
\right\}_{i=1}^{3}
\right].
\end{aligned}
\end{equation}
The worldline functional integral therefore reduces to ordinary Gaussian integrals, with Fourier-mode covariance
\begin{equation}
\begin{aligned}
 \int \mathscr{D}\mu_\alpha\; a_{ia}(m)a_{jb}(n) =\frac{\delta_{ij}\delta_{ab}\delta_{mn}}{m_\alpha\omega_n^2}.
\end{aligned}
\end{equation}
The particle path integral can then be evaluated using the Gaussian identities
\begin{equation}
\begin{aligned}
\int \mathscr{D}\mu_\alpha \; \xi_i(s)\xi_j(s')e^{\sum_l (ik'_l\xi_l(\tau')+ik_l\xi_l(\tau))}=\bigg(\delta_{ij}G_\alpha(s-s')+J_iJ'_j\bigg)e^{-\frac{1}{2}J},
\end{aligned}
\end{equation}
where
\begin{equation}
\begin{aligned}
J_i=ik'_iG_\alpha(s-\tau')+ik_iG_\alpha(s-\tau)\\
J'_j=ik'_jG_\alpha(s'-\tau')+ik_jG_\alpha(s'-\tau),\\
J=G_\alpha(0)\left(\mathbf{k}^2+\mathbf{k}'^{\,2}\right)
+2\,\mathbf{k}\cdot\mathbf{k}'\,G_\alpha(\tau-\tau')
\end{aligned}
\end{equation}
and
\begin{equation}
\begin{aligned}
\int \mathscr{D}\mu_\alpha \; \xi_i(s)e^{\sum_l (ik'_l\xi_l(\tau')+ik_l\xi_l(\tau))}=J_ie^{-\frac{1}{2}J},
\end{aligned}
\end{equation}
\begin{equation}
\begin{aligned}
\int \mathscr{D}\mu_\alpha \; e^{\sum_l (ik'_l\xi_l(\tau')+ik_l\xi_l(\tau))}=e^{-\frac{1}{2}J}.
\end{aligned}
\end{equation}
The particle Green function satisfies
\begin{equation}
\begin{aligned}
G_\alpha(\tau-\tau')=\frac{1}{m_\alpha}\bigg(\frac{1}{2\beta}|\tau-\tau'|^2-\frac{1}{2}|\tau-\tau'|+\frac{\beta}{12}\bigg),\\
\partial_\tau G_\alpha(\tau-\tau')=\frac{1}{m_\alpha}\bigg(\frac{1}{\beta}(\tau-\tau')-\frac{1}{2}\frac{|\tau-\tau'|}{(\tau-\tau')}\bigg),\quad\tau\neq\tau',\\ \partial_\tau G_\alpha(0)=0,\quad
\partial_\tau^2G_\alpha(\tau-\tau')=\frac{1}{m_\alpha}\bigg(\frac{1}{\beta}-\delta(\tau-\tau')\bigg).
\end{aligned}
\end{equation}
The relation $G'_\alpha(0)=0$ follows from the Fourier expansion. 
Substituting these Gaussian averages into the expansion of
$\Theta_\alpha$ to second order in the electromagnetic field gives the
response functions $I_\alpha$, $\chi_1$, and $\chi_2$ introduced in the main
text. The homogeneous dielectric background does not modify the free-particle
worldline measure; it enters through the bare electromagnetic operator and the
gauge-fixing term. Consequently, the same worldline correlators are retained,
whereas the electromagnetic eigenvalues acquire the replacement
\begin{equation}
k^2c^2+\omega_n^2
\longrightarrow
k^2c^2+\varepsilon_b\omega_n^2.
\end{equation}
Together with
$\kappa_{D,\alpha}^2\propto\varepsilon_b^{-1}$, this directly leads to the
factors $Z_1$--$Z_4$ given in the main text.

\section{Canonical partition function}\label{Canonical pf}

Let $\mathcal A$ be the plate area, $L$ the plate separation, and
$V=\mathcal A L$.
For a particle of species $\alpha$, define
\begin{equation}
w_\alpha[A;\mathbf x]
=
\int \mathscr{D}\mu_\alpha \; e^{-\frac{1}{\hbar}\int\limits_{0}^{\hbar\beta}d\tau \;\mathscr{U}_\alpha}
=
1+\Theta_\alpha[A;\mathbf x], \quad w_\alpha[0;\mathbf x]=1.
\end{equation}
In Maxwell--Boltzmann statistics, the canonical matter partition
function at a prescribed electromagnetic field is
\begin{equation}
Z_N[A]
=
\prod_\alpha
\frac{1}{N_\alpha!}
\left[
\frac{g_\alpha}{\Lambda_{\alpha}^3}
\int_V d^3x\,w_\alpha[A;\mathbf x]
\right]^{N_\alpha}.
\end{equation}
The normalization at zero field is
\begin{equation}
Z_N[0]
=
\prod_\alpha
\frac{1}{N_\alpha!}
\left(\frac{g_\alpha V}{\Lambda_{\alpha}^3}\right)^{N_\alpha}.
\end{equation}
Consequently,
\begin{equation}
\frac{Z_N[A]}{Z_N[0]}
=
\prod_\alpha
\left[
1+\frac{1}{V}\int_Vd^3x\,
\Theta_\alpha[A;\mathbf x]
\right]^{N_\alpha}.
\end{equation}
Taking the logarithm gives the exact canonical expression
\begin{equation}
\mathcal W[A]
\equiv
\ln\frac{Z_N[A]}{Z_N[0]}
=
\sum_\alpha N_\alpha
\ln\left[
1+\frac{1}{V}\int_Vd^3x\,
\Theta_\alpha[A;\mathbf x]
\right].
\end{equation}
We introduce the trajectory vertex
\begin{equation}
\mathcal V[A;\mathbf x,\boldsymbol\xi]
=
\int_0^{\beta \hbar}d\tau\,
\bigg(\varphi(\tau,\mathbf x+\boldsymbol\xi)-\dot{\boldsymbol\xi}\cdot\mathbf{A}(\tau,\mathbf x+\boldsymbol\xi)\bigg).
\end{equation}
Then
\begin{equation}
\begin{aligned}
\Theta_\alpha =  \int \mathscr{D}\mu_\alpha \; \bigg(e^{-\frac{iq_\alpha}{\hbar}\mathcal V}-1\bigg),
\end{aligned}
\end{equation}
The expansion
\begin{equation}
\Theta_\alpha
=
\Theta_\alpha^{(1)}+\Theta_\alpha^{(2)}+O(A^3)
\end{equation}
has coefficients
\begin{equation}
\Theta_\alpha^{(1)}
=
-\frac{iq_\alpha}{\hbar}
\langle\mathcal V\rangle_\alpha,
\qquad
\Theta_\alpha^{(2)}
=
-\frac{q_\alpha^2}{2\hbar^2}
\langle\mathcal V^2\rangle_\alpha.
\end{equation}
Using $\ln(1+x)=x-x^2/2+O(x^3)$ and $N_\alpha=n_\alpha V$, we obtain
\begin{equation}
\begin{aligned}
\mathcal W[A]
={}&
\sum_\alpha n_\alpha\int_Vd^3x\,\Theta_\alpha^{(1)}
+
\sum_\alpha n_\alpha\int_Vd^3x\,\Theta_\alpha^{(2)}
\\
&-
\frac12\sum_\alpha\frac{n_\alpha}{V}
\left[
\int_Vd^3x\,\Theta_\alpha^{(1)}
\right]^2
+O(A^3).
\end{aligned}
\end{equation}

\section{Source contribution to the partition function}\label{Source}
The average velocity vertex vanishes
\begin{equation}
\left\langle
\dot\xi_i(\tau)
e^{i\mathbf k\cdot\boldsymbol\xi(\tau)}
\right\rangle_\alpha=0.
\end{equation}
Then
\begin{equation}
\left\langle\dot{\boldsymbol\xi}\cdot\mathbf{A}(\tau,\mathbf x+\boldsymbol\xi)\right\rangle_\alpha=0.
\end{equation}

\begin{equation}
\int_Vd^3x\,\Theta_\alpha^{(1)} =-\frac{iq_\alpha}{\hbar}\int_Vd^3x\,\int_0^{\beta \hbar}d\tau\,
\langle\varphi(\tau,\mathbf x+\boldsymbol\xi)\rangle_\alpha.
\end{equation}

The expansion of the scalar potential is
\begin{equation}
\begin{aligned}
\varphi(\tau,\mathbf x_\parallel,z)
=
\frac{1}{\sqrt{\mathcal A}}
\sum_{\substack{l_1,\,l_2\in \mathbb{Z}\\ l\in \mathbb{N}}}
\sum_{(n,a)\in \mathcal{I}}\phi_a(n,l_1,l_2,l)\, \\ 
\times\; e^{i\mathbf k_\parallel\cdot\mathbf x_\parallel}
u_a(\tau,\beta,\omega_n)u_2(z,L,q_l),\quad \omega_n=\frac{2\pi n}{\hbar\beta},
\end{aligned}
\end{equation}
\begin{equation}
\phi_a(n,-l_1,-l_2,l)=\phi^{*}_a(n,l_1,l_2,l).
\end{equation}
Here 
\begin{equation}
\mathcal{I}
=
\{(0,1)\}
\cup
\{(n,a):n\in \mathbb{N},\ a=1,2\}.
\end{equation}
\begin{equation}
k_\parallel=(k_1,k_2),\quad k_j=\frac{2\pi l_j}{L_j},\quad j=1,2.
\end{equation}
The basis functions are
\begin{equation}
\begin{aligned}
u_1(\tau,\beta,\omega_n)=\sqrt{\frac2\beta}\cos (\omega_n \tau),\quad u_2(\tau,\beta,\omega_n)=\sqrt{\frac2\beta}\sin (\omega_n \tau),\quad n\neq0,\\
u_1(\tau,\beta,0)=\frac{1}{\sqrt{\beta}},\quad u_2(z,L,q_l)=\sqrt{\frac2L}\sin (q_l z),\quad q_l=\frac{\pi l}{L},
\end{aligned}
\end{equation}
The normalization conditions are
\begin{equation}
\frac1\hbar\int_0^{\hbar\beta}d\tau\,
u_a(\tau,\beta,\omega_n)
u_b(\tau,\beta,\omega_m)
=
\delta_{nm}\delta_{ab},
\qquad
(n,a),(m,b)\in\mathcal I.
\end{equation}
\begin{equation}
\int_0^Ldz\,u_2(z,L,q_l)u_2(z,L,q_{l'})=\delta_{ll'}.
\end{equation}
Since
\begin{equation}
\int_0^Ldz\,
\sqrt{\frac2L}\sin\frac{\pi lz}{L}
=
\begin{cases}
\displaystyle
\sqrt{\frac8L}\,\frac{1}{q_l},
& l\ \text{odd},\\[6pt]
0,& l\ \text{even},
\end{cases}
\end{equation}
only odd modes contribute to the integrated linear term. Using the identity
\begin{equation}
\left\langle
e^{i\mathbf k\cdot\boldsymbol\xi(\tau)}
\right\rangle_\alpha
=
\exp\left(-\frac{\lambda_{dB,\alpha}^2k^2}{24}\right),
\end{equation}
we obtain
\begin{equation}
\int_Vd^3x\,\Theta_\alpha^{(1)}
=
-iq_\alpha\sqrt{\beta\mathcal A}
\sum_{l=0}^{\infty}b_{\alpha l}a_l.
\end{equation}
Here we introduce
\begin{equation}
p_l=\frac{(2l+1)\pi}{L},
\qquad
a_l=\phi_1(0,0,0,2l+1),
\qquad
l=0,1,2,\ldots.
\end{equation}
\begin{equation}
E_\alpha(p)=
\exp\left(-\frac{\lambda_{dB,\alpha}^2p^2}{24}\right),
\qquad
b_{\alpha l}
=
\sqrt{\frac8L}\,\frac{E_\alpha(p_l)}{p_l}.
\end{equation}
The total linear term is
\begin{equation}
\sum_\alpha n_\alpha\int_Vd^3x\,\Theta_\alpha^{(1)}
=
-i\sqrt{\beta\mathcal A}\,
\boldsymbol\rho^T\boldsymbol a,
\end{equation}
where the source coefficients are
\begin{equation}
\rho_l=\sum_\alpha n_\alpha q_\alpha b_{\alpha l}, \quad \boldsymbol \rho^T\boldsymbol a = \sum_{l=0}^{\infty}\rho_la_l.
\end{equation}
We introduce the notation
\begin{equation}
\rho_l=\sum_\alpha n_\alpha q_{\alpha} b_{\alpha l}
=
\sqrt{\frac8L}\,\frac{\gamma(p_l)}{p_l},
\qquad
\gamma(p)=\sum_\alpha n_\alpha q_\alpha E_\alpha (p).
\end{equation}
The additional quadratic term is
\begin{equation}
\Delta\mathcal W^{(2)}
=
-\frac12\sum_\alpha\frac{n_\alpha}{V}
\left[
\int_Vd^3x\,\Theta_\alpha^{(1)}
\right]^2,
\end{equation}
Substituting the integrated linear term and using
$V=\mathcal A L$, we obtain
\begin{equation}
\begin{aligned}
\Delta\mathcal W^{(2)}
&=
-\frac12\sum_\alpha\frac{n_\alpha}{\mathcal A L}
\left[
-iq_\alpha\sqrt{\beta\mathcal A}
\sum_l b_{\alpha l}a_l
\right]^2
\\
&=
\frac{\beta}{2L}
\sum_\alpha n_\alpha q_\alpha^2
\left(\sum_l b_{\alpha l}a_l\right)^2.
\end{aligned}
\end{equation}
We introduce
\begin{equation}
\kappa_{D,\alpha}^2=\frac{\beta n_\alpha q_\alpha^2}{\varepsilon_0\varepsilon_b},
\qquad
U_{l\alpha}=\frac{\kappa_{D,\alpha}}{\sqrt L}\,b_{\alpha l}.
\end{equation}
Then
\begin{equation}
\Delta\mathcal W^{(2)}
=
\frac{\varepsilon_0\varepsilon_b}{2}
\boldsymbol a^TUU^T\boldsymbol a.
\end{equation}
The off-diagonal matrix arises only for modes of the scalar field $\varphi$ in the odd sector $p_l=\frac{(2l+1)\pi}{L}$, $\mathbf{k}_\parallel=0$, $n=0$. Then the action restricted to the source-coupled sector, through second order in the field, is
\begin{equation}
\frac{1}{\hbar}S_{0}=-\frac{\varepsilon_0\varepsilon_b}{2}
\boldsymbol a^T(K-UU^T)\boldsymbol a
-i\sqrt{\beta\mathcal A}\,
\boldsymbol\rho^T\boldsymbol a.
\end{equation}
where we define
\begin{equation}
\boldsymbol a^T(K-UU^T)\boldsymbol a = \sum_{l=0}^{\infty}\sum_{l'=0}^{\infty}  a_l(K_{ll'} - \sum_\alpha U_{l\alpha}U_{l'\alpha})a_{l'}.
\end{equation}
\begin{equation}
K_{ll'}= d_l\delta_{ll'},
\qquad
d_l=p_l^2+\chi_0(p_l),
\end{equation}
\begin{equation}
\chi_0(p)=\sum_\alpha\kappa_{D,\alpha}^2 I_\alpha(0,p).
\end{equation}
Thus, the canonical quadratic operator is
\begin{equation}
G_0=K-UU^T.
\end{equation}
For positive definite $G_0$, Gaussian integration yields
\begin{equation}
\begin{aligned}
&\int \prod_{l=0}^{\infty}da_l\,
\exp\left[
-\frac{\varepsilon_0\varepsilon_b}{2}\boldsymbol a^T G_0 \boldsymbol a
-i\sqrt{\beta\mathcal A}\,
\boldsymbol\rho^T\boldsymbol a
\right]
\\
&\qquad\propto
(\det G_0)^{-1/2}
\exp\left[
-\frac{\beta\mathcal A}{2\varepsilon_0\varepsilon_b}
\boldsymbol\rho^T G_0^{-1}\boldsymbol\rho
\right].
\end{aligned}
\end{equation}
The canonical source contribution is therefore
\begin{equation}
\ln Z_{4}
=
-\frac{\beta\mathcal A}{2\varepsilon_0\varepsilon_b}
\boldsymbol\rho^T(K-UU^T)^{-1}\boldsymbol\rho.
\end{equation}
We define a matrix in species space and a species-space vector
\begin{equation}
M=I-U^TK^{-1}U,
\qquad
\mathbf t=U^TK^{-1}\boldsymbol\rho.
\end{equation}
Using the identity
\begin{equation}
(K-UU^T)^{-1}
=
K^{-1}
+
K^{-1}UM^{-1}U^TK^{-1},
\end{equation}
we obtain
\begin{equation}
\ln Z_4
=
-\frac{\beta\mathcal A}{2\varepsilon_0\varepsilon_b}\bigg(
\boldsymbol\rho^TK^{-1}\boldsymbol\rho
+\mathbf t^TM^{-1}\mathbf t\bigg).
\end{equation}
The first term is determined by the identity
\begin{equation}
\begin{aligned}
\boldsymbol\rho^TK^{-1}\boldsymbol\rho
&=
\frac{8}{L}
\sum_{l=0}^{\infty}
\frac{\gamma^2(p_l)}{p_l^2d_l}.
\end{aligned}
\end{equation}
The corresponding free-energy surface density is
\begin{equation}
\frac{F_4}{\mathcal A}
=
-\frac{1}{\beta\mathcal A}\ln Z_4
=
\frac{1}{2\varepsilon_0\varepsilon_b}\bigg(\boldsymbol\rho^TK^{-1}\boldsymbol\rho
+
\mathbf t^TM^{-1}\mathbf t\bigg).
\end{equation}

The matrix elements and vector components are explicitly
\begin{equation}
M_{\alpha\beta}
=
\delta_{\alpha\beta}
-
\frac{8\kappa_{D,\alpha}\kappa_{D,\beta}}{L^2}
\sum_{l=0}^{\infty}
\frac{E_\alpha(p_l)E_\beta(p_l)}{p_l^2d_l},
\end{equation}
\begin{equation}
t_\alpha
=
\frac{8\kappa_{D,\alpha}}{L^{3/2}}
\sum_{l=0}^{\infty}
\frac{E_\alpha(p_l)\gamma(p_l)}{p_l^2d_l}.
\end{equation}
Thus a two-component plasma requires only the inversion of
a $2\times2$ matrix. Explicitly,
\begin{equation}
\mathbf t^TM^{-1}\mathbf t
=
\frac{
M_{22}t_1^2-2M_{12}t_1t_2+M_{11}t_2^2
}{
M_{11}M_{22}-M_{12}^2
}.
\end{equation}
To calculate $\det G_0$, we first restrict the number of modes to a finite cutoff $N$.
By Sylvester's determinant identity
\begin{equation}
\begin{aligned}
\frac{\det(K-UU^T)}{\det K}
&=
\det\nolimits_{N}
\left(I_{N}-K^{-1}UU^T\right)
\\
&=
\det\nolimits_2
\left(I_2-U^TK^{-1}U\right).
\end{aligned}
\end{equation}
For each finite mode cutoff, the determinant identity holds
exactly. Since the sums defining $M$ converge, removing the
cutoff gives the relative determinant
\begin{equation}
\frac{\det G_0}
     {\det K}=\lim_{N\to\infty}
\frac{\det G_0^{(N)}}
     {\det K^{(N)}}
=
\det M.
\end{equation}
\section{Sylvester's determinant identity}
Let $A$ be an $m\times n$ matrix and $B$ an $n\times m$ matrix. Then
\begin{equation}
\det (I_m+AB) = \det (I_n+BA),
\end{equation}
where $I_m$ is the identity matrix $m\times m$ and $I_n$ is the identity matrix $n\times n$.

\section{Casimir pressure in the ammonia background}
The Casimir pressure in the ammonia background is calculated from
\begin{equation}
P_{bg}(L)
=
-\frac{k_B T}{\pi}
\sum_{n=0}^{\infty}{}'
\int_{0}^{\infty} dq\,
q\,Q_n
\left[
\coth\left(LQ_n\right)-1
\right],
\end{equation}
where
\begin{equation}
Q_n
=
\sqrt{
q^2
+
\varepsilon_b\frac{\omega_n^2}{c^2}
},
\end{equation}
For the classical plasma, we use the identity
\begin{equation}
\begin{aligned}
P_{\mathrm{cl}}(L)
={}&
\frac{k_B T}{8\pi L^3}
\left[
\zeta(3)
-
\frac{1}{2}
\int_{2\kappa_D L}^{\infty}
dy\,
\frac{y^2}{e^y-1}
\right] + P_{bg}(L).
\end{aligned}
\end{equation}
The source contribution is
\begin{equation}\label{f_4}
f_4=\frac{1}{2\varepsilon_0\varepsilon_b} \bigg[\frac{8}{L}\sum_{l=0}^{\infty} h(p_{l}) +
\mathbf t^TM^{-1}\mathbf t\bigg]    
\end{equation}
where 
\begin{equation}
h(p)
=
\frac{\gamma^2(p)}
{p^2\left[p^2+\chi_0(p)\right]}.
\end{equation}
For an even function $h$, sampled at $p_l=\pi(2l+1)/L$, the Poisson summation formula gives
\begin{equation}
\frac{1}{L}
\sum_{l=0}^{\infty}
h\left(
\frac{\pi(2l+1)}{L}
\right)
=
\frac{1}{2\pi}
\int_0^\infty dp\,h(p)
+
\frac{1}{\pi}
\sum_{m=1}^{\infty}
(-1)^m
\int_0^\infty dp\,
h(p)\cos(mLp).
\end{equation}
Differentiating with respect to $L$ gives
\begin{equation}
-\frac{\partial}{\partial L}\bigg( \, \frac{1}{L}
\sum_{l=0}^{\infty}
h\left(p_l
\right)\bigg)
=
\frac{1}{\pi}
\sum_{m=1}^{\infty}
(-1)^m \, m
\int_0^\infty dp\,
p\,h(p)\sin(mLp).
\end{equation}
As a function of $p$, the oscillatory factor has period $2\pi/(mL)$, whereas $h(p)$ varies on the characteristic scale $1/\lambda_{dB,\alpha}$. For $L\gg\lambda_{dB,\alpha}$ the oscillatory period is therefore much shorter than the variation scale of $h(p)$, and the Fourier integrals are strongly suppressed by cancellation. Thus, within the model, the first term on the right-hand side of Eq.~(\ref{f_4}) gives only a small contribution to the Casimir force gradient.

A useful additional crossover parameter is
$\eta=\hbar c/(2\pi\sqrt{\varepsilon_b}k_BTL)$. Qualitatively,
$\eta\gg1$ corresponds to a regime dominated by zero-point (vacuum)
fluctuations, where the Mehra expansion is appropriate, whereas $\eta\ll1$
corresponds to a thermally dominated regime controlled primarily by the
$n=0$ term, where the Schwinger expansion is appropriate. In the
lithium--ammonia example $\eta\simeq0.45$--$0.56$; we therefore retain the
full sums over both Matsubara modes $n$ and normal-mode indices $l$.

\bibliography{Casimir_screening_quantum_Boltzmann_surface_force_refs}

@article{Lindhard1954,
  title = {On the Properties of a Gas of Charged Particles},
  author = {Lindhard, Jens},
  journal = {Kongelige Danske Videnskabernes Selskab, Matematisk-fysiske Meddelelser},
  volume = {28},
  number = {8},
  pages = {1--57},
  year = {1954}
}

@article{barouch1973interdependence,
  title = {On the interdependence of van der {W}aals and double layer forces},
  author = {Barouch, E. and Perram, J. W. and Smith, E. R.},
  journal = {Chemical Physics Letters},
  volume = {19},
  number = {1},
  pages = {131--133},
  year = {1973}
}

@article{gorelkin1973calculation,
  title = {Calculation of intermolecular interaction forces between bodies separated by a film of a strong electrolyte solution},
  author = {Gorelkin, V. N. and Smilga, V. P.},
  journal = {Soviet Journal of Experimental and Theoretical Physics},
  volume = {36},
  pages = {761},
  year = {1973}
}

@article{barnes1975statistical,
  title = {Statistical mechanics and {L}ifshitz theory for electrolytes. Part 1},
  author = {Barnes, C. J. and Davies, B.},
  journal = {Journal of the Chemical Society, Faraday Transactions 2},
  volume = {71},
  pages = {1667--1689},
  year = {1975}
}

@article{jancovici2004screening,
  title = {Screening of classical {C}asimir forces by electrolytes in semi-infinite geometries},
  author = {Jancovici, B. and {\v{S}}amaj, L.},
  journal = {Journal of Statistical Mechanics: Theory and Experiment},
  volume = {2004},
  number = {08},
  pages = {P08006},
  year = {2004}
}

@article{jancovici2005casimir,
  title = {Casimir force between two ideal-conductor walls revisited},
  author = {Jancovici, B. and {\v{S}}amaj, L.},
  journal = {Europhysics Letters},
  volume = {72},
  number = {1},
  pages = {35},
  year = {2005}
}

@article{buenzli2005microscopic,
  title = {Microscopic origin of universality in {C}asimir forces},
  author = {Buenzli, P. R. and Martin, P. A.},
  journal = {Journal of Statistical Physics},
  volume = {119},
  pages = {273--307},
  year = {2005}
}

@article{buenzli2008microscopic,
  title = {Microscopic theory of the {C}asimir force at thermal equilibrium: Large-separation asymptotics},
  author = {Buenzli, P. R. and Martin, P. A.},
  journal = {Physical Review E},
  volume = {77},
  number = {1},
  pages = {011114},
  year = {2008}
}

@article{maianeto2019scattering,
  title = {Scattering theory of the screened {C}asimir interaction in electrolytes},
  author = {Maia Neto, P. A. and Rosa, F. S. S. and Pires, L. B. and Moraes, A. B. and Canaguier-Durand, A. and Gu{\'e}rout, R. and Lambrecht, A. and Reynaud, S.},
  journal = {The European Physical Journal D},
  volume = {73},
  pages = {178},
  year = {2019},
  doi = {10.1140/epjd/e2019-100225-8}
}

@article{pires2021probing,
  title = {Probing the screening of the {C}asimir interaction with optical tweezers},
  author = {Pires, L. B. and Ether, D. S. and Spreng, B. and Ara{\'u}jo, G. R. S. and Decca, R. S. and Dutra, R. S. and Borges, M. and Rosa, F. S. S. and Ingold, G.-L. and Moura, M. J. B. and Frases, S. and Pontes, B. and Nussenzveig, H. M. and Reynaud, S. and Viana, N. B. and Maia Neto, P. A.},
  journal = {Physical Review Research},
  volume = {3},
  number = {3},
  pages = {033037},
  year = {2021},
  doi = {10.1103/PhysRevResearch.3.033037}
}

@article{budkovBrandyshev2025PRE,
  title = {First-principles theory of the {C}asimir screening effect},
  author = {Budkov, Yury A. and Brandyshev, Petr E.},
  journal = {Physical Review E},
  volume = {111},
  number = {4},
  pages = {044130},
  year = {2025},
  doi = {10.1103/PhysRevE.111.044130}
}

@article{brandyshevBudkov2025JMP,
  title = {Theory of {C}asimir forces: A unified approach using finite-temperature field theory},
  author = {Brandyshev, Petr E. and Budkov, Yury A.},
  journal = {Journal of Mathematical Physics},
  volume = {66},
  number = {12},
  pages = {123301},
  year = {2025},
  doi = {10.1063/5.0271407}
}

@article{Bivin,
  title = {Orbital magnetic susceptibility of electrons confined in a rectangular box},
  author = {Bivin, D. B. and McClure, J. W.},
  journal = {Phys. Rev. B},
  volume = {16},
  number = {2},
  pages = {762},
  year = {1977},
  doi = {10.1103/PhysRevB.16.762}
}

@article{cohenThompson1968,
  title = {The electronic and ionic structures of metal-ammonia solutions},
  author = {Cohen, Morrel H. and Thompson, James C.},
  journal = {Advances in Physics},
  volume = {17},
  number = {70},
  pages = {857--907},
  year = {1968},
  doi = {10.1080/00018736800101396}
}

@article{zurekEdwardsHoffmann2009,
  title = {A Molecular Perspective on Lithium--Ammonia Solutions},
  author = {Zurek, Eva and Edwards, Peter P. and Hoffmann, Roald},
  journal = {Angewandte Chemie International Edition},
  volume = {48},
  number = {44},
  pages = {8198--8232},
  year = {2009},
  doi = {10.1002/anie.200900373}
}

@article{chabanPrezhdo2016,
  title = {Electron Solvation in Liquid Ammonia: Lithium, Sodium, Magnesium, and Calcium as Electron Sources},
  author = {Chaban, Vitaly V. and Prezhdo, Oleg V.},
  journal = {The Journal of Physical Chemistry B},
  volume = {120},
  number = {9},
  pages = {2500--2506},
  year = {2016},
  doi = {10.1021/acs.jpcb.6b00412}
}

@article{somoanoThompson1970,
  title = {Optical Constants of Metal-Ammonia Solutions},
  author = {Somoano, R. B. and Thompson, J. C.},
  journal = {Physical Review A},
  volume = {1},
  pages = {376},
  year = {1970},
  doi = {10.1103/PhysRevA.1.376}
}

@article{perkynsKusalikPatey1986,
  title = {On the dielectric constant of liquid ammonia},
  author = {Perkyns, J. S. and Kusalik, P. G. and Patey, G. N.},
  journal = {Chemical Physics Letters},
  volume = {129},
  number = {3},
  pages = {258--261},
  year = {1986},
  doi = {10.1016/0009-2614(86)80207-X}
}

@article{fishMillerSmyth1958,
  title = {Microwave Absorption and Molecular Structure in Liquids. XXIV. The Dielectric Relaxation of Liquid Ammonia},
  author = {Fish, Kenneth and Miller, Robert C. and Smyth, Charles P.},
  journal = {The Journal of Chemical Physics},
  volume = {29},
  number = {4},
  pages = {745--746},
  year = {1958},
  doi = {10.1063/1.1744585}
}

@article{breitschwerdtSchmidt1970,
  title = {Permittivity and Dielectric Relaxations in Liquid-Ammonia Solutions},
  author = {Breitschwerdt, K. G. and Schmidt, W.},
  journal = {Zeitschrift f{\"u}r Naturforschung A},
  volume = {25},
  number = {10},
  pages = {1467--1471},
  year = {1970},
  doi = {10.1515/zna-1970-1018}
}

@article{kindtSchmuttenmaer1996,
  title = {Far-Infrared Dielectric Properties of Polar Liquids Probed by Femtosecond Terahertz Pulse Spectroscopy},
  author = {Kindt, J. T. and Schmuttenmaer, C. A.},
  journal = {The Journal of Physical Chemistry},
  volume = {100},
  number = {24},
  pages = {10373--10379},
  year = {1996},
  doi = {10.1021/jp960141g}
}

@article{liu2019precision,
  title = {Precision measurements of the gradient of the {C}asimir force between ultraclean metallic surfaces at larger separations},
  author = {Liu, Mingyue and Xu, Jun and Klimchitskaya, G. L. and Mostepanenko, V. M. and Mohideen, U.},
  journal = {Physical Review A},
  volume = {100},
  number = {5},
  pages = {052511},
  year = {2019},
  doi = {10.1103/PhysRevA.100.052511}
}

@book{israelachvili2011,
  author = {Israelachvili, Jacob N.},
  title = {Intermolecular and Surface Forces},
  edition = {3},
  publisher = {Academic Press},
  address = {Burlington, MA},
  year = {2011}
}

@book{derjaguinChuraevMuller1987,
  author = {Derjaguin, B. V. and Churaev, N. V. and Muller, V. M.},
  title = {Surface Forces},
  publisher = {Consultants Bureau},
  address = {New York},
  year = {1987},
  doi = {10.1007/978-1-4757-6639-4}
}

\end{document}